\documentclass{IEEEtran}

\usepackage{graphicx}
\graphicspath{ {./Figures/} }

\usepackage{graphics} 
\usepackage{epsfig} 
\usepackage{mathptmx} 
\usepackage{times} 
\usepackage{amsmath} 

\usepackage{booktabs}
\usepackage{tabularx}
\usepackage{siunitx} 

\usepackage[
backend=bibtex,
style=ieee,
sorting=none,
citestyle=numeric-comp
]{biblatex}
\AtBeginBibliography{\footnotesize}

\usepackage{amssymb}  
\usepackage{bm}
\usepackage{color}

\usepackage{graphicx}
\graphicspath{ {./Figures/} }

\usepackage{graphics} 

\usepackage{amssymb}
\usepackage{wrapfig}
\usepackage{multirow}
\usepackage{tabularx,booktabs}
\usepackage{algorithm} 
\usepackage{algpseudocode} 
\usepackage{chngcntr}

\newcolumntype{C}{>{\centering\arraybackslash}X} 
\usepackage{lipsum} 
\usepackage{mdframed}

\usepackage{float}
\usepackage{amsthm}
\newtheorem{theorem}{Theorem}

\ifCLASSOPTIONcompsoc
\else
\fi
\ifCLASSINFOpdf
\else
\fi
	\usepackage[table,xcdraw,dvipsnames]{xcolor} 
	\usepackage[colorlinks=true,
	linkcolor=blue,
	citecolor=blue,
	urlcolor=magenta]{hyperref}
	\usepackage[nameinlink,capitalize,noabbrev]{cleveref} 

\begin{document}
		%
		\title{Cross-Layer Reliability Assessment of Learning-Enabled UAV Navigation through Adversarial Edge-Case Discovery}
		%
		%
		%
		%
		
		\author{Deepak Kumar Panda
			\thanks{$^{1}$D. Kumar Panda is with the Faculty of Engineering and Applied Sciences,
				Cranfield University, MK43 0AL Cranfield, U.K
				{\tt\small Deepak.Panda@cranfield.ac.uk}}%
		}
		
		%
		%

	\markboth{Journal of \LaTeX\ Class Files,~Vol.~14, No.~8, August~2015}%
	{Shell \MakeLowercase{\textit{et al.}}: Bare Advanced Demo of IEEEtran.cls for IEEE Computer Society Journals}
	%



	\IEEEtitleabstractindextext{%
		\begin{abstract}
		Learning-enabled unmanned aerial vehicles (UAVs) integrate state estimation, learned navigation policies, and feedback control to translate sensor measurements into autonomous flight behaviour. In UAV navigation systems employing global navigation satellite system (GNSS) and inertial measurement units (IMU) sensor fusion, GNSS measurement integrity is commonly monitored through statistical consistency tests on estimator innovation residuals. However, existing studies predominantly investigate sensing integrity and policy robustness independently, overlooking adversarial edge cases arising from the propagation of statistically admissible sensing perturbations across autonomy layers. To address this gap, this paper presents an analytical and computational framework for discovering cross layer adversarial edge cases induced by GNSS spoofing. First, plant-model-independent bounds on estimation errors and innovation residuals under multi-rate sensor fusion are derived, requiring knowledge of the estimator structure rather than the full nonlinear UAV dynamics. Subsequently, sufficient conditions under which estimator-admissible perturbations degrade downstream navigation performance are established, linking innovation-level stealth to policy sensitivity. Guided by these analytical results, multi-objective Bayesian optimisation characterises the Pareto trade-off between mission degradation and estimator-level detectability. Closed-loop simulations demonstrate that the discovered adversarial edge cases reduce mission success from approximately 98\% to 28-29\%, while increasing median collision rates from 2\% to 70-72\%. Despite these consequences, window-level detection AUROC decreases from nearly 1.0 for conventional sinusoidal attacks to 0.50-0.56, while GRU-based episode-level alarm rates decrease from approximately 100\% to 10-15\%. These findings demonstrate that estimator-level statistical consistency does not guarantee mission-level reliability, highlighting the need for cross-layer reliability assessment of learning-enabled autonomous systems.
		\end{abstract}
		
		}

	\maketitle

	\IEEEdisplaynontitleabstractindextext

	%
	\IEEEpeerreviewmaketitle

	\ifCLASSOPTIONcompsoc
	\IEEEraisesectionheading{\section{Introduction}\label{sec:introduction}}
	\else
	\label{sec:introduction}
	\fi
	
	\section{Introduction}
	Modern unmanned aerial vehicles (UAVs) increasingly rely on learning-enabled autonomy for navigation in dynamic airspace \cite{PANDA2026124096}. In a typical navigation architecture, measurements from GNSS and inertial measurement unit (IMUs) are fused to estimate the vehicle state \cite{ye2023review}, which is used by a learned policy to generate commands for navigation in the lower-level control layers to accomplish the mission objectives \cite{marshall2021survey}. Therefore, the state-estimator serves as a critical interface between onboard sensing and autonomous decision- making, through which sensor perturbations propagate into downstream navigation decisions.
	
	The integrity of this navigation architecture is commonly assessed through monitoring mechanisms deployed at different stages in the navigation stack. At the estimator level, statistical consistency tests based on innovation residuals assess whether incoming measurement data is compatible with the predicted state and its uncertainty \cite{bar2001estimation, piche2016online}. At the state-space level, learning-based anomaly detectors characterize nominal navigation behaviour and identify deviations in their temporal evolution \cite{prashant2025guaranteeing, haider2023out}. However, these mechanisms provide layer-specific assessments without explicitly accounting for the propagation of statistically admissible sensing perturbations across interconnected estimation, decision-making, and control layers. Although prior studies have demonstrated that adversarial perturbations can evade estimator-level detection while degrading closed-loop performance \cite{bai2017data}, characterizing their propagation into estimation errors often requires knowledge of the underlying plant dynamics  \cite{wang2025quadformer, piche2016online}. This limits their applicability to practical UAV architectures, where the estimator structure may be accessible through open-source flight-control implementations \cite{baldi2022ardupilot}, but the nonlinear plant dynamics may not be fully known. Furthermore, how the resulting estimation errors influence learned navigation decisions and mission-level performance remains insufficiently characterized.
	
	In this paper, the interplay between estimator-level statistical detectability and downstream navigation performance is investigated through the systematic discovery of GNSS-spoofing-induced adversarial edge cases, without the knowledge of underlying UAV flight dynamics. First, estimation-error propagation across high-rate IMU-driven prediction and low-rate GNSS-based correction steps is characterized using the reduced-order propagation structure and bounded uncertainty, without requiring the complete nonlinear quadrotor dynamics. Then, sufficient statistical conditions are established linking estimator-admissible perturbations to downstream navigation performance degradation. These theoretical results are used to design a multi-objective optimization framework for discovering adversarial edge cases that expose the trade-off between estimator-level detectability and mission-level degradation.
	
	\subsection{Related Works}
	\subsubsection{Edge Cases for Safe Autonomy}
	Edge cases in learning-enabled autonomous systems lack a universally  accepted definition, although existing studies characterize them  primarily in terms of rarity and safety-criticality \cite{liang2026corner}. Early research focused on accelerated 
	estimation of rare safety-critical events in autonomous vehicles \cite{zhao2016accelerated,o2018scalable}. While accelerated evaluation efficiently quantifies rare-event risks, adaptive stress testing (AST) formulates failure discovery as a sequential decision-making problem, designing perturbations that drive the system towards failure states 
	\cite{koren2018adaptive,lee2020adaptive}. Complementary approaches, such as DeepXplore and DeepTest, employ coverage-guided testing to explore previously unobserved system behaviours and expose potential failures \cite{pei2017deepxplore,tian2018deeptest}. More recent methods 
	incorporate physical and environmental constraints into adversarial scenario generation to generate kinematically feasible perturbations \cite{hanselmann2022king, wang2021advsim}.
	
	However, kinematic feasibility does not ensure that perturbations remain undetected by the integrity-monitoring mechanisms deployed within autonomous systems. In this study adversarial edge cases refer to perturbations in sensor data that remain statistically admissible to prescribed integrity-monitoring mechanisms while inducing substantial degradation in downstream mission performance. 
	
	\subsection{Stealthy Attacks and Detectability Criteria in Autonomy}
	Estimator-level integrity monitoring commonly relies on the nominal statistical manifold of the innovation sequence to detect anomalous behavior. For Kalman-filter type estimators, the innovation residuals represent the difference between the observed and predicted measurements, while their covariance quantifies the associated uncertainty. Statistical consistency tests of these innovation residuals assess whether they remain compatible with the stochastic model of the estimator \cite{teixeira2011cyber}. Hence, a stealthy adversary can manipulate sensor measurements while constraining the distributions of the resulting innovation residuals to remain within the acceptance region of the anomaly detector. 
	
	Early studies on innovation residuals based attack detection demonstrated that structured false-data injection (FDI) attacks could compromise power-system state estimation while evading conventional residual tests \cite{liu2011false}. Mo \textit{et al.} subsequently investigated attacks against Kalman-filter-based estimation, while Pasqualetti \textit{et al.} and Vamvoudakis \textit{et al.} examined structural and detection-theoretic conditions for adversarial detectability \cite{mo2010false,pasqualetti2013attack,vamvoudakis2014detection}. Teixeira \textit{et al.} formalized adversarial capabilities in terms of model knowledge and access to sensing and actuation channels, demonstrating how resource constraints can influence both attack feasibility and physical consequences \cite{teixeira2015secure}. Related studies have formulated attack strategies that degrade estimation performance while satisfying statistical stealth constraints on the compromised innovations \cite{zhou2023optimal}.
	
	Beyond estimator-level statistical admissibility, a key challenge is to establish the conditions under which adversarial perturbations degrade closed-loop performance. Existing studies have investigated the maximization of estimation errors subject to stealth constraints, highlighting the trade-off between attack detectability and impact 
	\cite{teixeira2015secure,zhang2022design,urbina2016limiting}. However, in learning-enabled autonomy, the operational consequences of estimation errors depend on the sensitivity of the learned state-to-action mapping and the resulting closed-loop behaviour. Characterizing adversarial impact therefore requires consideration of policy sensitivity beyond the magnitude of estimation errors alone.
	
	Statistical admissibility at the estimation layer does not necessarily  imply behavioural consistency over extended temporal horizons. Classical one-class methods identify deviations from nominal data distributions \cite{scholkopf2001estimating,tax2004support}, while deep 
	one-class and energy-based approaches extend anomaly detection to high-dimensional representations \cite{ruff2018deep,ruff2021unifying,liu2020energy}. In autonomous 
	navigation, sequential and trajectory-level detectors capture anomalies arising from the temporal evolution of individually plausible states \cite{jiao2023learning,sun2026driving}. Although these methods complement estimator-level monitoring, their effectiveness depends on the nominal data distribution, temporal representation, and operating conditions. Estimator-level and trajectory-level detectability therefore constitute distinct criteria for assessing adversarial stealth.
	
	These limitations motivate a cross-layer analysis linking estimator-level statistical admissibility, policy sensitivity, and mission-level performance without requiring complete knowledge of the nonlinear UAV dynamics.
	\begin{figure*}[!t]
		\centering
		\includegraphics[width=0.95\textwidth]{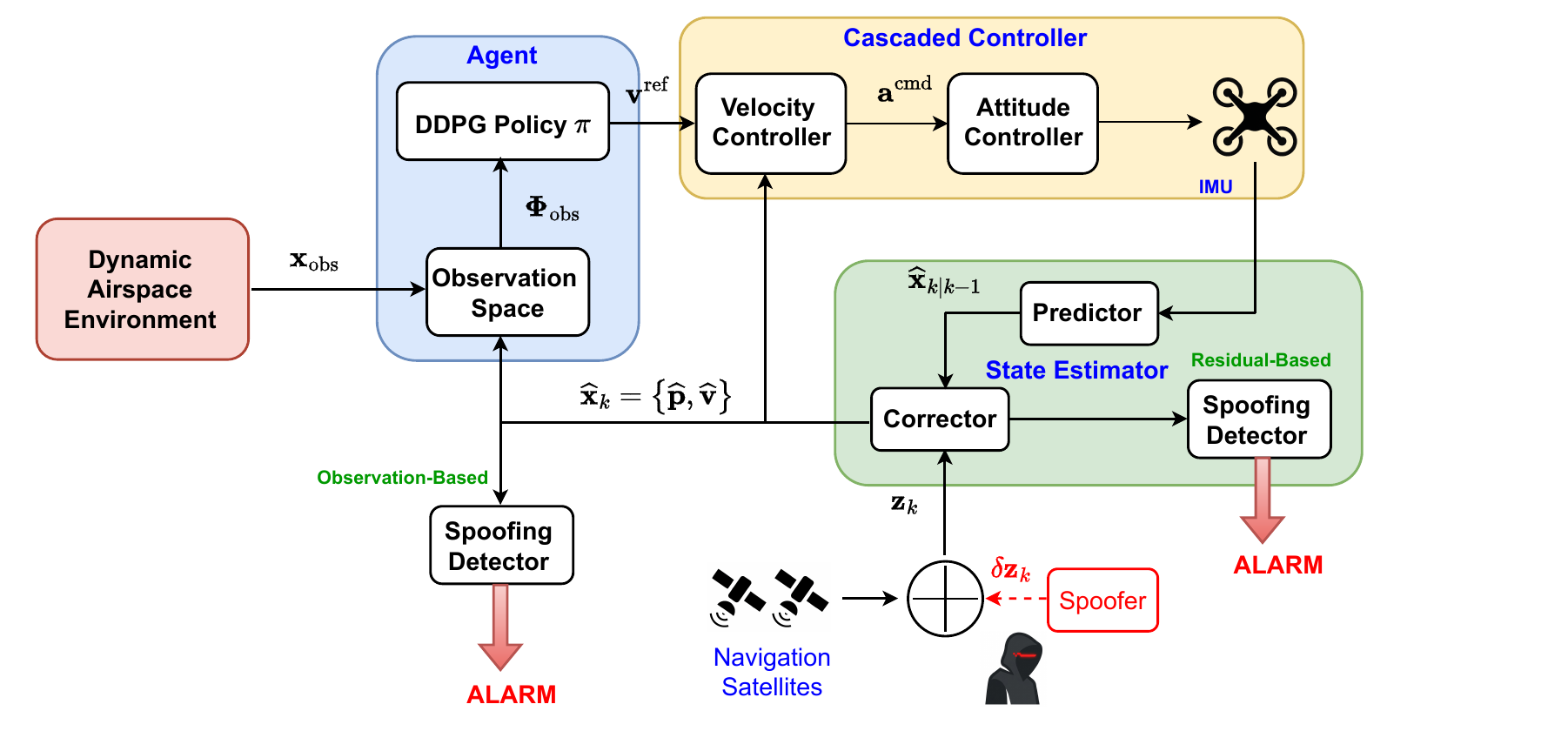}
		\caption{Schematic of the proposed learning-enabled UAV navigation framework integrating state estimation with observation- and residual-based GNSS spoofing detection in dynamic airspace environments.}
		\label{figure1_schematic}
	\end{figure*}
	
	\subsection{Research Gap}
	Despite advances in adversarially robust navigation, stealthy attack analysis, and learning-based anomaly detection, several challenges remain in understanding the relationship between estimator-level statistical detectability and downstream navigation consequences. \textit{First}, existing model-dependent analyses often rely on knowledge of the underlying system dynamics to characterize the propagation of adversarial perturbations into estimation errors. This raises the question of how innovation-residual-based detectability conditions can be established using the estimator's prediction-correction structure and bounded propagation uncertainty, without the knowledge of underlying nonlinear plant model. \textit{Second}, estimator-level statistical admissibility constrains the behaviour of measurement innovations but does not, by itself, determine the consequences of perturbation-induced deviations in the estimated state for downstream navigation. Establishing the conditions under which these deviations propagate through the policy's state-to-action mapping and result in closed-loop navigation degradation is therefore a central theoretical challenge. \textit{Third}, although the trade-off between attack detectability and system performance has been investigated, its implications for the systematic discovery of navigation edge cases under estimator-level statistical constraints remain insufficiently understood. In particular, the extent to which such edge cases remain undetected by monitoring mechanisms requires investigation. Together, these challenges motivate a policy-aware framework that connects estimator-level statistical admissibility downstream navigation consequences without requiring the knowledge of complete nonlinear UAV dynamics.
	\subsection{Contributions}
	
	In order to address the research gaps, the paper develops a cross-layer framework for identifying the adversarial edge cases in learning enabled navigation. The principal contribution of the paper are threefold.
	
	\begin{enumerate}
		\item \textbf{Plant-model-independent characterization of innovation-based attack detectability:} The multi-rate estimation-error and innovation-residual bounds are derived using known reduced-order estimator dynamics and bounded propagation uncertainty, without requiring the complete nonlinear UAV plant model. These results support the characterization of statistical acceptance regions for GNSS spoofing perturbations under innovation-based consistency monitoring.
		
		\item \textbf{Sufficient conditions for policy degradation under estimator-admissible perturbations:} A cross-layer policy-sensitivity relationship is established connecting measurement perturbations with estimator corrections and downstream mission return. The analysis identifies sufficient local conditions under which a non-zero perturbation remains admissible to the estimator-level monitor while inducing positive expected mission degradation.
		
		\item \textbf{Policy-aware discovery and cross-layer evaluation of adversarial edge cases: } GNSS spoofing attack discovery is formulated as a multi-objective optimization problem that maximizes navigation-performance degradation while minimizing estimator-level detectability. Multi-objective Bayesian optimization is used to identify Pareto-efficient attacks, whose consequences are subsequently assessed through mission outcomes and residual-based and learning-based state-space detectors.
	\end{enumerate}
	\section{System Description}
	Figure~\ref{figure1_schematic} illustrates the proposed learning-enabled UAV navigation architecture integrating the DDPG agent, cascaded controller, state estimator, equipped with multi-layer detection mechanisms. Observation features $\Phi_{\mathrm{obs}}$ are supplied to the DDPG policy $\pi$ to generate the action which becomes the reference velocity $\mathbf{v}^{\mathrm{ref}}$ for the quadrotor's cascaded controller. The velocity reference command $\mathbf{v}^{\mathrm{ref}}$ is converted into the acceleration command $\mathbf{a}^{\mathrm{cmd}}$ through the velocity control loop, followed by attitude-control loop which is converted into relevant motor thrust command for the quadrotor. In parallel, a state estimator employing a predictor-corrector structure uses inertial sensor data to compute the prior state estimate $\hat{\mathbf{x}}_{k|k-1}$, which is updated using the satellite navigation measurement $\mathbf{z}_k$ to obtain the corrected state estimate $\hat{\mathbf{x}}_k=\{\hat{\mathbf{p}}_k,\hat{\mathbf{v}}_k\}$. The resulting corrected navigation state, comprising the estimated position and velocity along with data from the airspace environment, forms the agent's observation for downstream decision-making. In order to detect the GNSS spoofing, two complementary mechanisms are incorporated: an observation-based detector that monitors the estimated navigation state $\hat{\mathbf{x}}_k$, and a residual-based detector within the state estimator that evaluates the consistency of the navigation measurements with the predicted state. As illustrated, a spoofer introduces an adversarial perturbation $\delta\mathbf{z}_k$ into the navigation-satellite measurements, while the two detection pathways independently generate alarms upon identifying anomalous navigation information.
	\subsection{Quadrotor Model and Cascaded Control System}
	\label{subsec:quad_control}
	The UAV is represented by the nonlinear rigid-body quadrotor model described in \cite{panda2026belief}. The vehicle state is
	\begin{equation}\label{eq:1}
		\mathbf{x}
		= \begin{bmatrix}
			\mathbf{p}^{T} & \mathbf{v}^{T} & \boldsymbol{\eta}^{T} & \boldsymbol{\omega}^{T}
		\end{bmatrix}^{T},
	\end{equation}
	where $\mathbf{p}=[x,y,z]^T$ and $\mathbf{v}=[v_x,v_y,v_z]^T$ denote the inertial-frame position and velocity, respectively, $\boldsymbol{\eta}=[\phi,\vartheta,\psi]^T$ contains the roll, pitch, and yaw angles, and $\boldsymbol{\omega}=[p,q,r]^T$ denotes the body-frame angular rates of the quadrotor. The translational and rotational dynamics, including the thrust and body-torque inputs, follow the formulation shown in \cite{panda2026belief} and are therefore not repeated here. As shown in Fig.~\ref{figure1_schematic}, the RL policy does not directly command the quadrotor actuators. Instead, it generates a reference velocity $\mathbf{v}^{\mathrm{ref}}_k$, which is tracked by a cascaded controller. The outer velocity loop converts the velocity-tracking error into a commanded acceleration,
	\begin{equation} \label{eq:2}
		\mathbf{a}^{\mathrm{cmd}} = \mathbf{K}^c_{p} \left(\mathbf{v}^{\mathrm{ref}}-\hat{\mathbf{v}}\right) + \mathbf{K}^c_{i} \int
		\left(\mathbf{v}^{\mathrm{ref}}-\hat{\mathbf{v}}\right)dt + \mathbf{K}^c_{d} \frac{d}{dt} \left(\mathbf{v}^{\mathrm{ref}}-\hat{\mathbf{v}}\right),
	\end{equation}
	where $\hat{\mathbf{v}}$ is supplied by the state estimator. The commanded acceleration is subsequently mapped to thrust and attitude references, which are tracked by the faster inner attitude loop. The detailed rigid-body dynamics, acceleration-to-attitude mapping, and inner-loop control formulation are provided in our previous work \cite{panda2026belief}. This hierarchical structure isolates high-level navigation decisions from the fast vehicle stabilization dynamics.
	\subsection{Estimator Model}\label{subsec:estimator}
	The RL agent and the cascaded controller rely on navigation states estimated by an extended Kalman filter (EKF), based on the methodology described in our previous work \cite{panda2026belief}, inspired by open-source flight control architectures \cite{fornasier2024equivariant}. The EKF propagates the estimated state using high-frequency inertial measurements, while lower-frequency GNSS position measurements are incorporated to correct the predicted state. For the present study, the estimator state is
	\begin{equation}\label{eq:3}
		\hat{\mathbf{x}}^{e}_{k} = \begin{bmatrix}
			\hat{\mathbf{p}}_{k}^{T} &	\hat{\mathbf{v}}_{k}^{T}
		\end{bmatrix}^{T},
	\end{equation}
	with the discrete-time propagation model
	\begin{equation} \label{eq:4}
		\hat{\mathbf{x}}^{e}_{k \mid k-1} = \mathbf{F}^{\mathrm{tran}} \hat{\mathbf{x}}^{e}_{k-1 \mid k-1} + \mathbf{B}\mathbf{a}^{m}_{k},
	\end{equation}
	where $\mathbf{a}^{m}_{k}$ denotes the IMU-derived acceleration measurement. At a GNSS update, the position measurement $\mathbf{z}_{k}$ is incorporated through the conventional EKF correction
	\begin{equation}\label{eq:5}
	\begin{gathered}
		\mathbf{y}_{k} = \mathbf{z}_{k} - \mathbf{H}\hat{\mathbf{x}}^{e}_{k \mid k-1} \\
		\mathbf{S}_{k} = \mathbf{H}\mathbf{P}_{k|k-1}\mathbf{H}^{T} + \mathbf{R}^{\mathrm{cov}}, \\
		\hat{\mathbf{x}}^{e}_{k \mid k} = \hat{\mathbf{x}}^{e}_{k \mid k-1} + \mathbf{K}_{k}\mathbf{y}_{k}
	\end{gathered}
	\end{equation}
	where $\mathbf{y}_{k}$ and $\mathbf{S}_{k}$ are the innovation and its covariance, respectively. Further details of the estimator prediction-correction structure and its integration with the quadrotor flight-control loop are provided in \cite{panda2026belief}. The resulting position and velocity estimates, $\hat{\mathbf{p}}_{k}$ and $\hat{\mathbf{v}}_{k}$, provide the navigation feedback to the controller and constitute part of the observation space available to the RL agent. Consequently, the navigation policy acts on the estimated, rather than the ground-truth vehicle state.
	\subsection{RL Agent and Dynamic Airspace Environment} \label{subsec:rl_environment} 
	The RL-based navigation problem is formulated as a continuous-state, continuous-action Markov decision process (MDP), \(\mathcal{M}=(\mathcal{S},\mathcal{A},\mathcal{P},r,\gamma)\), where \(\mathcal{S}\) and \(\mathcal{A}\) represent the state and action spaces, \(\mathcal{P}\) denotes the state-transition probability kernel, \(r\) is the reward function, and \(\gamma\in[0,1)\) is the discount factor. The deterministic policy \(\pi_{\theta_v} \), parameterized by \(\theta_v\), maps the observation vector \(\boldsymbol{\Phi}_k\) to a continuous navigation action \(\mathbf{a}_k=\pi_{\theta_v} (\boldsymbol{\Phi}_k)\).
	\subsubsection{Observation Space}
	The policy observation combines the UAV state, obtained from the estimator described in Section \ref{subsec:estimator}, with relative kinematic data associated with the closest dynamic obstacle in the airspace environment. Let $\mathbf{c}_{k}^{\mathrm{obs}}$ and $\mathbf{v}_{k}^{\mathrm{obs}}$ denote the position and velocity of the closest obstacle respectively, and let $R_{\mathrm{obs}}$ denote its effective collision radius similar to the one described in our previous work \cite{PANDA2026124096}. The relative obstacle feature is defined as
	\begin{equation}\label{eq:6}
		\mathbf{s}^{\mathrm{obs}}_{k} = \left(\mathbf{c}_{k}^{\mathrm{obs}} - \hat{\mathbf{p}}_{k} \right) \frac{d_{k}^{\mathrm{obs}}-R_{\mathrm{obs}}}{d_{k}^{\mathrm{obs}}},
	\end{equation}
	where $d_{k}^{\mathrm{obs}} = \left\| \mathbf{c}_{k}^{\mathrm{obs}} - \hat{\mathbf{p}}_{k} \right\|_{2}.$ The goal-relative displacement is $\mathbf{s}^{g}_{k} = \mathbf{p}^{g}-\hat{\mathbf{p}}_{k},$ with $\mathbf{p}^{g}$ denoting the goal position. The complete observation space is therefore defined by
	\begin{equation}\label{eq:7}
		\boldsymbol{\Phi}_{k} = \begin{bmatrix}
			(\mathbf{s}^{\mathrm{obs}}_{k})^{T} & (\mathbf{s}^{g}_{k})^{T} & (\mathbf{v}^{\mathrm{obs}}_{k})^{T} & \hat{\mathbf{v}}_{k}^{T}
			\end{bmatrix}^{T}
		\in\mathbb{R}^{12}.
	\end{equation}
	\subsubsection{Action Space}
	The policy produces a normalized three-dimensional continuous action, $\mathbf{\grave{a}}_{k} = \begin{bmatrix} \grave{a}_{x,k} & \grave{a}_{y,k} & \grave{a}_{z,k} \end{bmatrix}^{T}, \mathbf{\grave{a}}_{k}\in[-1,1]^{3}.$ The action is scaled to obtain the reference translational velocity, $\mathbf{v}^{\mathrm{ref}}_{k} = v_{\max}\mathbf{a}_{k},$ with the vertical component additionally constrained by the prescribed vertical-velocity limit. The reference velocity is passed to the cascaded controller described in Section \ref{subsec:quad_control}. 
	\subsubsection{Dynamic Airspace and Transition Model}
	The considered Class G airspace contains autonomous aerial vehicles whose positions and velocities evolve over time according to a set of heterogeneous motion patterns \cite{PANDA2026124096}. At the beginning of an episode, obstacle motion modes and phase offsets are sampled to generate variations in obstacle trajectories. Within one RL transition, $\mathbf{v}^{\mathrm{ref}}_{k}$ is held as the high-level command while the faster velocity-control, attitude-control, vehicle-dynamics, IMU-propagation, and GNSS-correction processes evolve at their respective update rates. 
	\subsubsection{Reward and Episode Termination}
	The reward function balances progress toward the navigation goal with dynamic obstacle avoidance while promoting smooth trajectories. Let $d^{g}_{k} = \left\| \hat{\mathbf{p}}_{k}-\mathbf{p}^{g} \right\|_{2}$ denote the estimated distance to the goal and let $c_{k}$ denote the true geometric clearance from the nearest obstacle boundary. The instantaneous reward is expressed as
	\begin{equation}\label{eq:8}
		\begin{split}
			r_{k} ={}&
			-\frac{d^{g}_{k}}{d^{g}_{0}}
			+\lambda_{p}
			\left(d^{g}_{k-1}-d^{g}_{k}\right)
			+r_{\mathrm{goal}}\mathbb{I}_{g}\\
			&-\lambda_{s}
			\left[d_{\mathrm{safe}}-c_{k}\right]_{+}
			-\lambda_{v}
			\left\|\mathbf{v}^{\mathrm{ref}}_{k}\right\|_{2}^{2}
			-\lambda_{\eta}
			\left\|\boldsymbol{\eta}^{p}_{k}\right\|_{2}^{2}
			-r_{\mathrm{col}}\mathbb{I}_{c},
		\end{split}
	\end{equation}
	Here, $[x]_{+}=\max(0,x)$ denotes the positive-part operator, while $\mathbb{I}_{g}$ and $\mathbb{I}_{c}$ are binary indicators of goal attainment and collision, respectively. The normalized distance term $-d_k^g/d_0^g$ penalizes the remaining distance to the goal, whereas $\lambda_p(d_{k-1}^g-d_k^g)$ rewards incremental progress, with $\lambda_p>0$ determining its relative contribution. Successful goal attainment is further reinforced by the reward $r_{\mathrm{goal}}\mathbb{I}_{g}$. To encourage obstacle avoidance, the term $-\lambda_s[d_{\mathrm{safe}}-c_k]_+$ imposes a soft penalty whenever the clearance from the closest dynamic obstacle falls below the prescribed safety threshold $d_{\mathrm{safe}}$. The quadratic penalty $-\lambda_v\|\mathbf{v}_k^{\mathrm{ref}}\|_2^2$ regularizes the reference velocity magnitude, while $-\lambda_\eta\|\boldsymbol{\eta}^{p}_k\|_2^2$, where $\boldsymbol{\eta}^{p}_k=\mathbf{v}_k^{\mathrm{ref}}-\mathbf{v}_{k-1}^{\mathrm{ref}}$, penalizes successive changes in velocity commands to promote smoother commanded motion. Finally, $r_{\mathrm{col}}>0$ determines the magnitude of the additional collision penalty $-r_{\mathrm{col}}\mathbb{I}_{c}$. The resulting objective of the policy is \[ \pi^{*} = \arg\max_{\pi} \mathbb{E}_{\pi} \left[ \sum_{k=0}^{T} \gamma^{k}r_{k}. \right] \] The deterministic policy is learned using deep deterministic policy gradient (DDPG), which is suitable for the continuous velocity-command action space considered here.
	
	\section{Attack Detectability Criteria} \label{sec:detectability}
	We characterize the conditions under which GNSS spoofing remains statistically admissible to the estimator-level monitor in Fig.~\ref{figure1_schematic}. The analysis is formulated at the estimation layer, using the predicted state, innovation covariance, and Kalman gain, while considering the data from inertial and GNSS sensors. This distinction is important because the estimator described in Section~\ref{subsec:estimator} propagates a reduced-order position-velocity state using inertial measurements, whereas the underlying vehicle evolves according to the nonlinear rigid-body dynamics.
	\subsection{Residual Innovation Bound} \label{subsec:detectability_assumptions}
	Since IMU measurements are available at a higher frequency than GNSS measurements, the indices $k$ and $i$ are used to denote the state propagation and GNSS correction instants, in the estimator explained in subsection \ref{subsec:estimator} respectively. Let
	\[
	\boldsymbol{\xi}_k = [ \mathbf{p}_k^T, \mathbf{v}_k^T]^T, \qquad \hat{\boldsymbol{\xi}}_k = [\hat{\mathbf{p}}_k^T, \hat{\mathbf{v}}_k^T]^T
	\]
	denote the true and estimated translational navigation states, respectively. Superscripts $\left( - \right)$ and $\left( + \right)$ denote the quantities immediately before and after a GNSS correction, hence $\mathbf{e}^{-}_i = \boldsymbol{\xi}_i - \hat{\boldsymbol{\xi}}_i^{-}$ and $\mathbf{e}^{+}_i = \boldsymbol{\xi}_i - \hat{\boldsymbol{\xi}}_i^{+}$ correspond to the prior and posterior estimation errors. The UAV acceleration and its associated measurement error are assumed to satisfy the following engineering bounds,
	\begin{equation}\label{eq:9}
		\|\mathbf{a}_{k}\|_{2}\leq \bar{\mathbf{a}}<\infty, \qquad 
		\|\boldsymbol{\eta}^{a}_{k}\|_{2}\leq \bar{\eta}_{a}<\infty.
	\end{equation}
	At each GNSS update instant \(i\), we assume that the Kalman gain and the posterior error-update are  uniformly bounded as,
	\begin{equation}\label{eq:10}
	\|\mathbf{K}_{i}\|_{2}\leq\bar{K}, \qquad \|\mathbf{I}-\mathbf{K}_{i}\mathbf{H}\|_{2} \leq\bar{\gamma},
	\end{equation}
	for finite constants \(\bar{K}\) and \(\bar{\gamma}\). In addition, the innovation covariance is given by
	\begin{equation}\label{eq:11}
	\mathbf{S}_{i} = \mathbf{H}\mathbf{P}_{i}^{-}\mathbf{H}^{T} +\mathbf{R},
	\end{equation}
	while satisfying $ \mathbf{S}_{i}\succ0$ at every GNSS update instant. Let \(\nu_i\) denote the nominal GNSS measurement noise and $\delta_i$ be the additive perturbation due to spoofing attack. We define the combined measurement disturbance as $d_i = \nu_i + \delta_i$. The innovation residual at the $i$-th GNSS update, after spoofing attack, as
	\begin{equation} \label{eq:12}
		\tilde{\mathbf{y}}_{i} = \mathbf{H}\mathbf{e}_{i}^{-} + \boldsymbol{\nu}_{i} + \boldsymbol{\delta}_{i} = \mathbf{H}\mathbf{e}_{i}^{-} + \mathbf{d}_{i},
	\end{equation}
	Considering the norms of the relation given in~\eqref{eq:12}, we can write
	\begin{equation}\label{eq:13}
		\|\tilde{\mathbf{y}}_{i}\|_{2} \leq \|\mathbf{e}_{i}^{-}\|_{2} + \|\mathbf{d}_{i}\|_{2},
	\end{equation}
	where \(\mathbf{H}=[\mathbf{I}_{3}\ \mathbf{0}]\). If we consider the bound on the disturbance as $\|\mathbf d_i\|_2 =\|\boldsymbol\nu_i+\boldsymbol\delta_i\|_2 \leq\bar d<\infty.$ Thus, the innovation residual magnitude is directly bounded by the maximum estimation error and maximum GNSS disturbance magnitude.
	\subsection{Reduced-Order Prediction-Error Bound} \label{subsec:prediction_bound}
	As indicated by (\ref{eq:13}), the deterministic innovation bound depends on an upper bound on the prior estimation error $\mathbf{e}_i^-$. Deriving this error bound directly from the nonlinear quadrotor dynamics \cite{panda2026belief} would introduce an explicit dependence on the plant model into the detectability analysis. To avoid this, we restrict the analysis to the reduced translational state $\boldsymbol{\xi}_k$ used at the estimator interface. The corresponding discrete-time state-transition matrix $\mathbf{A}$ and IMU input mapping matrix $\mathbf{B}_T$, to describe the evolution of the translational state $\xi$ over each EKF prediction interval \cite{panda2026belief} as follows:
	\begin{equation} \label{eq:14}
		\boldsymbol{\xi}_{k+1}= \mathbf{A} \boldsymbol{\xi}_k + \mathbf{B}_T \mathbf{a}_k. 
	\end{equation}
	The estimator utilizes the acceleration measurement from IMU sensors $ \mathbf{a}_k^m= \mathbf{a}_k+ \boldsymbol{\eta}_k^a$ and the input matrix $\mathbf{B}_E$. Subtracting the estimator propagation we get,
	\begin{equation} \label{eq:15}
		\mathbf{e}_{k+1}^- = \mathbf{A} \mathbf{e}_k^+ + \tilde{\mathbf{w}}_k, 
	\end{equation}
	where $\tilde{\mathbf{w}}_k \triangleq (\mathbf{B}_T - \mathbf{B}_E) \mathbf{a}_k - \mathbf{B}_E \boldsymbol{\eta}_k^a$
	is the effective propagated disturbance. The first term $(\mathbf{B}_T - \mathbf{B}_E) \mathbf{a}_k$ captures the mismatch between physical and estimator propagation, while the second term $ \mathbf{B}_E \boldsymbol{\eta}_k^a$ captures the error due to  accelerometer measurement. 
	\subsection{Effective Propagation-Disturbance Bound}
	In order to obtain a finite bound on the prediction error, we first obtain the bound for propagated disturbance $\tilde{\mathbf w}_{k}$ in (\ref{eq:15}). Taking the induced Euclidean norm and applying the triangle inequality we obtain
	\begin{equation} \label{eq:17}
		\begin{aligned}
			\left\|\tilde{\mathbf w}_{k}\right\|_{2} &= \left\| (\mathbf B_T-\mathbf B_E)\mathbf a_k -
			\mathbf B_E\boldsymbol{\eta}^{a}_{k} \right\|_{2} \\[2mm]
			&\leq \left\|\mathbf B_T-\mathbf B_E\right\|_{2} \left\|\mathbf a_k\right\|_{2} + \left\|\mathbf B_E\right\|_{2} \left\|\boldsymbol{\eta}^{a}_{k}\right\|_{2}.
		\end{aligned}
	\end{equation}
	Utilizing the physical bound  on UAV acceleration and the measurement error bound in (\ref{eq:9}) along with the propagation disturbance bound in (\ref{eq:17}), we can write
	\begin{equation}\label{eq:18}
		\boxed{\left\|\tilde{\mathbf w}_{k}\right\|_{2} \leq \bar w_{\mathrm{eff}}}
	\end{equation}
	where \[ \bar w_{\mathrm{eff}} = \left\|\mathbf B_T-\mathbf B_E\right\|_{2}\bar a + \left\|\mathbf B_E\right\|_{2}\bar\eta_a. \]
	Computing the respective norms, we obtain
	\begin{equation} \label{eq:19}
		\begin{gathered}
			\left\|\mathbf B_T-\mathbf B_E\right\|_{2} = \frac{1}{2}\Delta t^{2}, \\
			\left\|\mathbf B_E\right\|_{2} = \Delta t \sqrt{1+\frac{\Delta t^{2}}{4}}.
		\end{gathered}
	\end{equation}
	Considering the matrix norms in (\ref{eq:19}), the closed form compact expression of bounded propagation disturbance, which becomes part of reduced-order prediction dynamics, is given by
	\begin{equation} \label{eq:20}
		\boxed{\bar w_{\mathrm{eff}} = \frac{1}{2}\Delta t^{2}\bar a + \Delta t \sqrt{1+\frac{\Delta t^{2}}{4}}\, \bar\eta_a}.
	\end{equation}
	\subsection{Multi-Rate Estimation-Error Bound}
	This subsection investigates how bounded disturbances propagate and accumulate over successive inertial prediction steps between GNSS corrections, using the bound derived in (\ref{eq:20}), to establish the resulting innovation bound for the estimator-level integrity monitor. 
	Let there be $L$ estimator propagation steps between two consecutive GNSS updates, which is finite and fixed throughout the operation. Hence, repeated  one-step prediction-error dynamics (\ref{eq:15}) over $L$ propagation steps gives:
	\begin{equation}\label{eq:21}
		\mathbf e_i^{-} = \mathbf A^{L}\mathbf e_{i-1}^{+} + \sum_{j=0}^{L-1} \mathbf A^{L-1-j}\tilde{\mathbf w}_{i,j},
	\end{equation}
	where $\mathbf e_{i-1}^{+}$ is the posterior estimation error following the GNSS correction and $\tilde{\mathbf w}_{i,j}$ denotes the effective propagation disturbance at the \(j\)-th propagation step. For convenience, let us define the accumulated propagation factor as
	\[
	\Gamma_L \triangleq \sum_{j=0}^{L-1} \left\|\mathbf A^{j}\right\|_2 .
	\]
	We now establish the relationship between multi-rate estimation-error and innovation bounds as per the following theorem.
	\begin{theorem}\label{th:1} 
	Considering the bounds on the Kalman gain, posterior error-update matrix and propagation disturbance in (\ref{eq:10}) and (\ref{eq:20}) respectively, we define the contraction coefficient as 
	\begin{equation}\label{eq:22}
		\rho\triangleq\sup_i \|(\mathbf{I}-\mathbf{K}_i \mathbf{H})\mathbf{A}^L\|_2,
	\end{equation}
	while considering $\rho < 1$. Then the posterior estimation error is asymptotically bounded as
	\begin{equation} \label{eq:23}
		\limsup_{i\to\infty} \left\|\mathbf{e}_{i}^{+}\right\|_{2} \leq \frac{\bar{\gamma}\Gamma_{L}\bar{w}_{\mathrm{eff}} + \bar{K}\bar{d}} {1-\rho},
	\end{equation}
	and the corresponding prior estimation error satisfies 
	\begin{equation} \label{eq:24}
		\limsup_{i\to\infty} \left\|\mathbf{e}_{i}^{-}\right\|_{2} \leq \left\|\mathbf{A}^{L}\right\|_{2}
		\frac{\bar{\gamma}\Gamma_{L}\bar{w}_{\mathrm{eff}} + \bar{K}\bar{d}}{1-\rho} + \Gamma_{L}\bar{w}_{\mathrm{eff}}.
	\end{equation}
	Consequently, the deterministic bound on the innovation residual of the estimator is given by
	\begin{equation} \label{eq:25}
	\begin{aligned}
		\limsup_{i\to\infty} \left\|\tilde{\mathbf{y}}_{i}\right\|_{2} &\leq \left\|\mathbf{A}^{L}\right\|_{2}
			\frac{\bar{\gamma}\Gamma_{L}\bar{w}_{\mathrm{eff}}+ \bar{K}\bar{d}}{1-\rho} + \Gamma_{L}\bar{w}_{\mathrm{eff}}+ \bar{d} \\
			&\triangleq B_y
	\end{aligned}
	\end{equation}
	\end{theorem}
	\begin{IEEEproof}
		Taking the Euclidean norm of both sides of (\ref{eq:21}), applying triangle inequality, and invoking the propagation-disturbance bound in (\ref{eq:18}), we obtain
		\begin{equation}\label{eq:26}
			\begin{aligned}
				\left\|\mathbf e_i^{-}\right\|_2 &\leq \left\|\mathbf A^{L}\right\|_2 \left\|\mathbf e_{i-1}^{+}\right\|_2 +
				\sum_{j=0}^{L-1} \left\|\mathbf A^{L-1-j}\right\|_2 \left\|\tilde{\mathbf w}_{i,j}\right\|_2
				\\[1mm]
				&\leq \left\|\mathbf A^{L}\right\|_2 \left\|\mathbf e_{i-1}^{+}\right\|_2 + \Gamma_L\bar w_{\mathrm{eff}} .
			\end{aligned}
		\end{equation}
		At the $i$th GNSS correction, we can write the translational state from the estimator correction equation as,
		\begin{equation}\label{eq:27}
			\hat{\boldsymbol{\xi}}_i^{+} = \hat{\boldsymbol{\xi}}_i^{-} + \mathbf K_i\tilde{\mathbf y}_i .	
		\end{equation}
		Substituting the attacked innovation residual in (\ref{eq:12}) into the estimator correction relation (\ref{eq:27}), while using the posterior estimation-error definition \(\mathbf e_i^+=\boldsymbol{\xi}_i-\hat{\boldsymbol{\xi}}_i^+\), we obtain
		\begin{equation} \label{eq:28}
			\mathbf e_i^{+} = (\mathbf I-\mathbf K_i\mathbf H)\mathbf e_i^{-} - \mathbf K_i\mathbf d_i.
		\end{equation}
		Substituting~\eqref{eq:21} into~\eqref{eq:28}, we get
		\[
				\mathbf{e}_i^+
				={} (\mathbf{I}-\mathbf{K}_i\mathbf{H})
				\mathbf{A}^{L}\mathbf{e}_{i-1}^{+}\\
				+(\mathbf{I}-\mathbf{K}_i\mathbf{H})
				\sum_{j=0}^{L-1}
				\mathbf{A}^{L-1-j}\tilde{\mathbf{w}}_{i,j}\\
				-\mathbf{K}_i\mathbf{d}_i.
		\]
		Taking the Euclidean norm and applying the triangle
		inequality and submultiplicativity of the induced
		matrix norm yields
		\begin{equation}\label{eq:29}
			\begin{aligned}
				\|\mathbf{e}_i^{+}\|_2
				&\leq{}
				\|(\mathbf{I}-\mathbf{K}_i\mathbf{H})
				\mathbf{A}^{L}\|_2
				\|\mathbf{e}_{i-1}^{+}\|_2 \\
				&+\|\mathbf{I}-\mathbf{K}_i\mathbf{H}\|_2
				\quad\times\sum_{j=0}^{L-1}
				\|\mathbf{A}^{L-1-j}\|_2
				\|\tilde{\mathbf{w}}_{i,j}\|_2
				+\|\mathbf{K}_i\|_2\|\mathbf{d}_i\|_2.
			\end{aligned}
		\end{equation}
		Combining (\ref{eq:29}) with the bounds established in the previous subsections and the contraction coefficient defined in (\ref{eq:22}), we obtain
		\begin{equation} \label{eq:30}
			\left\|\mathbf e_i^{+}\right\|_2 \leq \rho
			\left\|\mathbf e_{i-1}^{+}\right\|_2 + \bar\gamma\Gamma_L\bar w_{\mathrm{eff}} + \bar K\bar d ,
		\end{equation}
		If we define \[c \triangleq \bar\gamma\Gamma_L\bar w_{\mathrm{eff}} + \bar K\bar d , \]
		The recursive relation in (\ref{eq:30}) can be written as
		\begin{equation} \label{eq:31}
			\left\|\mathbf e_i^{+}\right\|_2 \leq \rho \left\|\mathbf e_{i-1}^{+}\right\|_2+c .
		\end{equation}
		Considering (\ref{eq:31}) across multiple GNSS update steps, 
		\begin{equation} \label{eq:32}
			\left\|\mathbf e_i^{+}\right\|_2 \leq \rho^{i} \left\|\mathbf e_0^{+}\right\|_2 + c \sum_{\ell=0}^{i-1}\rho^\ell .
		\end{equation}
		As $\rho < 1$, we can define the limit for (\ref{eq:32}) for a large number of update steps as,
		\begin{equation} \label{eq:33}
			\rho^i\left\|\mathbf e_0^{+}\right\|_2 \rightarrow0, \qquad
			\sum_{\ell=0}^{i-1}\rho^\ell \rightarrow \frac{1}{1-\rho},
		\end{equation}
		Utilizing the limiting conditions in (\ref{eq:33}), and the expression of $c$ in (\ref{eq:32}), we obtain the relation given in (\ref{eq:23}). Similarly, when we substitute the relation given in (\ref{eq:23}) into (\ref{eq:26}), we obtain the relation (\ref{eq:24}).
		Using the prior estimation-error bound in (\ref{eq:13}) and the innovation-residual relation in (\ref{eq:20}), together with the bounded measurement disturbance (\ref{eq:30}), we obtain the asymptotic upper bound on the innovation-residual magnitude given in (\ref{eq:25})
	\end{IEEEproof}	
	Theorem \ref{th:1} establishes a deterministic asymptotic upper bound on the innovation-residual magnitude under bounded propagation uncertainty and additive measurement disturbances. To relate this bound to estimator-level statistical consistency monitoring, we consider the normalized innovation squared (NIS) at $i-$the GNSS correction, defined as,
	\begin{equation}\label{eq:34}
		\mathbf{D}_i \triangleq \tilde{\mathbf y}_i^{T} \mathbf S_i^{-1} \tilde{\mathbf y}_i ,
	\end{equation}
	To establish the nominal acceptance region, the innovation-residual bound in (\ref{eq:25}) is evaluated under attack-free conditions, i.e., \(\boldsymbol\delta_i=\mathbf 0\) and \(\bar d=\bar\nu\), yielding the nominal innovation bound \(B_y^{\mathrm{nom}}\).
	Consequently, if \(\mathbf S_i\succeq\underline s\mathbf I\), with \(\underline s>0\), we can derive the NIS bound as
	\begin{equation} \label{eq:35}
	\begin{aligned}
			\limsup_{i\rightarrow\infty}\mathbf{D}_i
			&\leq \frac{1}{s}
			\limsup_{i\rightarrow\infty}
			\|\tilde{\mathbf{y}}_i\|_2^2\\
			&\leq \frac{B_y^2}{s}
			\triangleq \tau_D.
	\end{aligned}
	\end{equation}
	Here, \(B_y\) is the deterministic innovation-residual bound derived in (\ref{eq:25}), and \(\tau_D\) is the corresponding deterministic NIS rejection threshold. Unlike conventional calibration based on $\chi^2$ distribution, the NIS rejection threshold \(\tau_D\) is established from the deterministic innovation-residual bound derived in Theorem \ref{th:1} under nominal operating conditions. Accordingly, a GNSS perturbation is estimator-admissible at the \(i\)-th correction instant when
	\begin{equation} \label{eq:36}
		\mathbf{D}_i = \left(\mathbf{H} \mathbf{e}^-_i + \nu_i + \delta_i \right)^T \mathbf{S}^{-1} \left(\mathbf{H} \mathbf{e}^-_i + \nu_i + \delta_i \right) \leq \tau_D
	\end{equation} 
	where $\delta_i$ represents a set of estimator-admissible GNSS perturbations at $i-$th correction defined as,
	\[
		\mathcal E_i\triangleq
		\left\{
		\boldsymbol\delta_i\in\mathbb R^3:
		\mathbf{D}_i(\boldsymbol\delta_i)\leq\tau_D
		\right\}.
	\]
	Hence, of the nominal innovation satisfies \(D_i(\mathbf 0)<\tau_D\), the origin lies in the interior of \(\mathcal E_i\). Under the stochastic operation conditions, expected NIS rejection probability at the $i-$th correction instant is defined as,
	\begin{equation}\label{stat:det}
	\begin{aligned}
		p_{\mathrm{det},i}(\boldsymbol\delta_i)
		&\triangleq
		\mathbb E\left[
		\mathbb I[D_i(\boldsymbol\delta_i)>\tau_D]
		\right]\\
		&=\Pr\left(D_i(\boldsymbol\delta_i)>\tau_D\right).
	\end{aligned}
	\end{equation}
	Here, the expectation is taken over the stochastic operating conditions, and \(p_{\mathrm{det},i}(\mathbf 0)\) denotes the nominal NIS rejection probability.
	\subsection{Policy Sensitivity under Estimator-Stealthy Perturbations} \label{subsec:detectability_implications}
	The NIS acceptance criterion (\ref{eq:36}), characterizes the statistical admissibility of GNSS perturbations at the estimator level but does not necessarily guarantee downstream navigation integrity. To quantify the sensitivity of the learned navigation policy to estimator-admissible perturbations, we define the cross-layer policy sensitivity matrix as follows:
	\begin{equation} \label{eq:37}
			\mathbf M_i=\mathbf J_{\pi,i}\mathbf G_i\mathbf K_i,
	\end{equation}
	where \(\mathbf{J}_{\pi,i}=\partial\pi_{\theta_v}/\partial\boldsymbol{\Phi}\) and \(\mathbf{G}_i=\partial g/\partial\hat{\boldsymbol{\xi}}\) are evaluated at the nominal corrected estimate. Here, \(g\) denotes the observation mapping defined by (\ref{eq:6}) and (\ref{eq:7}), which maps the estimated translational state to the policy observation vector. This formulation enables us to establish the existence of estimator-admissible perturbations that degrade navigation performance and characterize the impact-detectability trade-off.
	\begin{theorem} \label{thm2}
	Let \(\mathcal R_i(\mathbf a_i)\) denote the expected mission reward return when the action at correction instant \(i\) is set to \(\mathbf a_i\), with the subsequent evolution governed by existing closed-loop navigation architecture. Suppose that the nominal innovation lies strictly within the thresholds in (\ref{stat:det}), the estimator and policy mappings are locally differentiable with respect to the nominal action \(\mathbf a_i^0\), while considering
	\begin{equation} \label{eq:38}
		\mathbf M_i^T
		\nabla_{\mathbf a}\mathcal R_i(\mathbf a_i^0)
		\neq\mathbf0.
	\end{equation}
	Then there exists a sufficiently small, nonzero perturbation that produces strictly positive expected reward degradation. Furthermore, if the nominal expected NIS rejection rate defined in (\ref{eq:36}) is strictly below a prescribed budget \(\varepsilon\) and the expected detection rate is locally continuous under the perturbation, the attack can be chosen to satisfy
	\begin{equation}\label{eq:39}
		f_{\mathrm{imp}}(\boldsymbol\delta_i)>0,
		\qquad
		f_{\mathrm{det}}(\boldsymbol\delta_i)\leq\varepsilon.
	\end{equation}
	\end{theorem}
	\begin{IEEEproof}
		From the estimator corrector equation in (\ref{eq:5}), a GNSS perturbation applied at the same prior estimate induces a change  in the corrected state estimate given by \(\Delta\widehat{\boldsymbol\xi}_i=\mathbf K_i\boldsymbol\delta_i\). Applying first-order expansions to the observation mapping and navigation policy gives
		\begin{equation}\label{eq:40}
			\Delta\mathbf a_i = \mathbf M_i\boldsymbol\delta_i + o(\|\boldsymbol\delta_i\|_2).
		\end{equation}
		Consequently, the first-order variation in the expected remaining mission return associated with the perturbation-induced change in the navigation satisfies
		\begin{equation} \label{eq:41}
		\begin{aligned}
			\Delta\mathcal R_i
			&=
			\nabla_{\mathbf a}\mathcal R_i(\mathbf a_i^0)^T
			\mathbf M_i\boldsymbol\delta_i
			+o(\|\boldsymbol\delta_i\|_2)\\
			&=
			\mathbf b_i^T\boldsymbol\delta_i
			+o(\|\boldsymbol\delta_i\|_2),
		\end{aligned}
		\end{equation}
		where \(\mathbf b_i=\mathbf M_i^T\nabla_{\mathbf a}\mathcal R_i(\mathbf a_i^0)\neq\mathbf0\).
		Choosing the perturbation direction $\boldsymbol\delta_i = -\tilde{t}\frac{\mathbf b_i}{\|\mathbf b_i\|_2},
		\quad \tilde{t}>0$ and substituting it into (\ref{eq:41}) yields
		\[
		\Delta\mathcal R_i
		=
		-\tilde{t}\|\mathbf b_i\|_2 + o(\tilde{t})<0
		\]
		for sufficiently small \(\tilde{t}\). Since the nominal innovation lies strictly inside the NIS acceptance region, a sufficiently small \(\tilde{t}\) also ensures \(\boldsymbol\delta_i\in\mathcal E_i\). Hence, the expected reward degradation satisfies
		\[
		f_{\mathrm{imp}}(\boldsymbol\delta_i)
		=
		-\Delta\mathcal R_i>0.
		\]
		Finally, continuity of the expected detection rate and the strict nominal detection margin imply that sufficiently small \(\tilde{t}\) preserves \(f_{\mathrm{det}}(\boldsymbol\delta_i)\leq\varepsilon\). Thus, a nonzero GNSS perturbation simultaneously induces positive expected reward degradation and satisfies the prescribed estimator-detection constraint.
	\end{IEEEproof}
	Theorem \ref{thm2} establishes a local relationship between estimator-induced policy sensitivity and the expected  mission rewards to navigation actions. While the analysis characterizes the policy-mediated pathway, the closed-loop evaluation captures the combined effects of measurement perturbations across estimation, decision-making, and control. This motivates the MOO formulation in the next section.
	\section{Multi-Objective Identification of Estimator-Stealthy Edge Cases} \label{sec:moo}
	As we have seen in Section \ref{sec:detectability}, the statistical admissibility of GNSS perturbations is characterized, while sufficient local conditions for downstream navigation degradation is established as per Theorem \ref{thm2}. Building on these results, we formulate adversarial edge-case discovery as a multi-objective optimization problem to investigate the trade-off between navigation degradation and estimator-level detectability.
	The spoofer considered in the analysis is similar to the ones done in our previous work \cite{panda2026curr, panda2026belief, PANDA2026124096}. This trade-off depends on perturbation characteristics, estimator uncertainty, and downstream policy sensitivity. Rather than imposing fixed objective weights, the proposed formulation characterizes the achievable trade-off through Pareto-efficient solutions within the considered parameter space.
	\subsection{Spoofing Policy Design}
	Let the adversarial strategy be parameterized by $\theta \in \Theta = [ -\theta_{\max}, \theta_{\max} ]^d$, where $d$ denotes the number of attack-policy parameters. At the $i$-th GNSS measurement update, the attacker constructs a feature vector,
	\[
		\boldsymbol{\phi}_i = \begin{bmatrix}
			\phi_{i,1} & \phi_{i,2} & \cdots & \phi_{i,m}
		\end{bmatrix}^{T},
	\]
	from the information available to the attack policy. The corresponding GNSS perturbation is generated according to bounded state-dependent mapping. 
	\begin{equation} \label{eq:42}
		\boldsymbol{\delta}_i(\boldsymbol{\theta}) = \delta_{\max} \tanh \left( \mathbf W_{\boldsymbol{\theta}} \boldsymbol{\phi}_i \right),
	\end{equation}
	where $\mathbf W_{\boldsymbol{\theta}}$ is determined by $\boldsymbol{\theta}$ and $\delta_{\max}$ bounds the injected perturbation. The estimator then receives the following measurement as, 
	\begin{equation} \label{eq:43}
		\tilde{\mathbf z}_i = \mathbf z_i + \boldsymbol{\delta}_i(\boldsymbol{\theta}).
	\end{equation}
	 During optimization, the estimator structure and victim navigation policy remain fixed.
	\subsection{Impact-Detectability Multi-Objective Formulation}
	For a given candidate attack policy \(\boldsymbol{\theta}\), let the expected discounted closed-loop return be defined as
	\begin{equation}\label{eq:44}
	 J(\boldsymbol{\theta}) = \mathbb E \left[\sum_{k=0}^{T-1} \gamma^k r_k \,\mid \, \boldsymbol{\theta} \right],
	\end{equation}
	where the expectation is taken over the stochastic closed-loop episodic simulations. Let \(J_{\mathrm{clean}}\) denote the corresponding expected return without spoofing. We quantify attack impact by the degradation relative to nominal operation, 
	\begin{equation}\label{eq:45}
		f_{\mathrm{imp}}(\boldsymbol{\theta}) = J_{\mathrm{clean}} - J(\boldsymbol{\theta}).	
	\end{equation}
	Thus, $f_{\mathrm{imp}}(\boldsymbol{\theta})>0$ indicates degradation relative to the unattacked system, with larger values corresponding to larger degradation. The second objective measures exposure to the estimator-level consistency monitor developed in Section \ref{sec:detectability}. For each valid GNSS measurement update \(i\), let us define 
	\begin{equation}\label{eq:46}
		v_i(\boldsymbol{\theta}) = \mathbb I \left[ \mathbf{D}_i(\boldsymbol{\theta})>\tau_{D} \right].
	\end{equation}
 	If \(\mathcal I(\boldsymbol{\theta})\) denotes the collection of innovation residuals within nominal manifold obtained during evaluation, the empirical estimator-detection rate is
	\begin{equation}\label{eq:47}
		f_{\mathrm{det}}(\boldsymbol{\theta}) = \frac{1}{|\mathcal I(\boldsymbol{\theta})|} \sum_{i\in\mathcal I(\boldsymbol{\theta})} v_i(\boldsymbol{\theta}).
	\end{equation} 
	The adversarial edge-case search is therefore formulated as
	\begin{equation}\label{eq:48}
		\underset{\boldsymbol{\theta}\in\Theta} {\operatorname{maximize}} \quad \mathbf F(\boldsymbol{\theta}) = \begin{bmatrix} f_{\mathrm{imp}}(\boldsymbol{\theta})\\[1mm] -f_{\mathrm{det}}(\boldsymbol{\theta})
		\end{bmatrix}
	\end{equation}
	which simultaneously optimizes to cause maximum impact on navigation with low estimator detectability. For two attack policies \(\boldsymbol{\theta}_a\) and \(\boldsymbol{\theta}_b\), \(\boldsymbol{\theta}_a\) dominates \(\boldsymbol{\theta}_b\), denoted $ \boldsymbol{\theta}_a \succ \boldsymbol{\theta}_b, $ if $f_{\mathrm{imp}}(\boldsymbol{\theta}_a) \geq f_{\mathrm{imp}}(\boldsymbol{\theta}_b)$ and $f_{\mathrm{det}}(\boldsymbol{\theta}_a) \leq f_{\mathrm{det}}(\boldsymbol{\theta}_b),$ with at least one strict inequality. The Pareto-optimal attack set can be represented as
	\begin{equation} \label{eq:49}
		\Theta^{\star} = \left\{ \boldsymbol{\theta}\in\Theta: \nexists\, \boldsymbol{\theta}' \in\Theta \text{ such that } \boldsymbol{\theta}'\succ\boldsymbol{\theta} \right\},
	\end{equation}
	with the corresponding impact-detectability frontier given by
	\begin{equation}\label{eq:50}
		\mathcal P^{\star} = \left\{ \mathbf F(\boldsymbol{\theta}): \boldsymbol{\theta}\in\Theta^{\star} \right\}.
	\end{equation}
	\subsection{Sample-Efficient Pareto Search}
	Direct evaluation of \(\mathbf{F}(\boldsymbol{\theta})\) requires repeated closed-loop simulations integrating UAV dynamics, state estimation, the navigation policy, and the dynamic environment under GNSS spoofing. Consequently, \(\mathbf{F}:\Theta\rightarrow\mathbb{R}^{2}\) is treated as an expensive stochastic black-box function with unavailable analytical gradients. This motivates the use of multi-objective Bayesian optimization (MOBO), which employs a probabilistic surrogate to guide evaluations efficiently under a limited computational budget. For numerical optimization, the bounded parameters are normalized \cite{daulton2021parallel} as follows:
	\begin{equation}\label{eq:51}
		\mathbf x = \frac{1}{2} \left(\frac{\boldsymbol{\theta}}{\theta_{\max}} + \mathbf 1 \right) \in[0,1]^{\tilde{d}} ,
	\end{equation}
	with the inverse transformation $\boldsymbol{\theta} = \theta_{\max} (2\mathbf x-\mathbf 1).$ An initial set of $N_0$ attack parameters, $\mathcal D_0 = \left\{ (\mathbf x_n,\mathbf y_n) \right\}_{n=1}^{N_0},$ is evaluated where $ \mathbf y_n = \begin{bmatrix} f_{\mathrm{imp}}(\boldsymbol{\theta}_n)\\ -f_{\mathrm{det}}(\boldsymbol{\theta}_n) \end{bmatrix}.$ A space-filling Sobol design is used for initialization with the zero-attack configuration. At iteration \(t\), the accumulated observations $\mathcal D_t = \left\{ (\mathbf x_n,\mathbf y_n) \right\}_{n=1}^{N_t}$ are used to construct a probabilistic surrogate for the unknown objective vector. If we denote the surrogate by \(\mathbf F(\mathbf x)\), the posterior may be written compactly as,
	\begin{equation} \label{eq:52}
		\mathbf{F}(\cdot)\mid\mathcal{D}_t \sim \mathcal{GP}\left(
		\boldsymbol{\mu}_t(\cdot), \boldsymbol{\Sigma}_t(\cdot,\cdot)
		\right).
	\end{equation}
	where \(\mathcal{GP}\) denotes the Gaussian process model. The posterior mean predicts the objective values, while the posterior covariance quantifies uncertainty, enabling a balance between exploration and exploitation. Let \(\mathcal{P}_t\) denote the current non-dominated objective set and \(\mathbf{r}\in\mathbb{R}^{2}\) a dominated reference point. The hypervolume \(\mathrm{HV}(\mathcal{P}_t;\mathbf{r})\) quantifies the region of objective space dominated by \(\mathcal{P}_t\) and bounded by \(\mathbf{r}\), providing a measure of Pareto-front approximation quality. For a new objective vector \(\mathbf{y}\), the hypervolume improvement is defined as
	\begin{equation}\label{eq:53}
		\Delta \mathrm{HV} (\mathbf y;\mathcal P_t,\mathbf r) = \mathrm{HV} (\mathcal P_t\cup\{\mathbf y\};\mathbf r) - \mathrm{HV} (\mathcal P_t;\mathbf r).
	\end{equation}
	For a batch of $q$ candidate attack parameters, $\mathbf X_q = \{\mathbf x^{(1)},\ldots,\mathbf x^{(q)}\},$ the qEHVI acquisition function evaluates the expected increase in dominated hypervolume under the current surrogate posterior,
	\begin{equation}\label{eq:54}
		\alpha_{\mathrm{qEHVI}} (\mathbf X_q) = \mathbb E_{\mathbf F(\mathbf X_q)\mid\mathcal D_t} \left[\Delta \mathrm{HV} \left( \mathbf F(\mathbf X_q); \mathcal P_t,\mathbf r \right) \right].
	\end{equation}
	The next attack parameters are therefore selected as
	\begin{equation}\label{eq:55}
		\mathbf X_q^{\star} = \underset{\mathbf X_q\subset[0,1]^d} {\operatorname{arg\,max}} \; \alpha_{\mathrm{qEHVI}}(\mathbf X_q).
	\end{equation}
	Each selected candidate \(\boldsymbol{\theta}^{(j)}\) is evaluated using repeated independent closed-loop rollouts to obtain empirical estimates 
	\begin{equation}\label{eq:56}
		\hat{\mathbf F} (\boldsymbol{\theta}^{(j)}) = \begin{bmatrix} \hat f_{\mathrm{imp}}(\boldsymbol{\theta}^{(j)})\\ -\hat f_{\mathrm{det}}(\boldsymbol{\theta}^{(j)})
		\end{bmatrix}.
	\end{equation}
	These observations are appended to \(\mathcal D_t\), the surrogate is updated, and (\ref{eq:53})-(\ref{eq:54}) are repeated until the search budget is exhausted. As the objectives $\hat{\mathbf F} (\boldsymbol{\theta}^{(j)})$ are estimated from finite stochastic rollouts, a candidate may therefore appear non-dominated during the search due to the stochastic variation rather than underlying expected performance which is a well-recognized issue in MOBO \cite{daulton2021parallel}. Accordingly, after completion of the Bayesian search, the provisional non-dominated set $
	\widehat{\Theta}^{\star} = \left\{ \boldsymbol{\theta}_n:
	n\in\operatorname{ND}_{\mathrm{idx}} \left( (\widehat{\mathbf F}(\boldsymbol{\theta}_j))_{j=1}^{N}
	\right) \right\}$ is re-evaluated using a larger independent rollout evaluation budget $M_R$. Non-dominated sorting is then repeated using these re-evaluated objectives to obtain the confirmed empirical Pareto set,
	\begin{equation}\label{eq:57}
		\Theta_{\mathrm{conf}}^\star = \left\{ \boldsymbol{\theta}_n^p: n\in\operatorname{ND}_{\mathrm{idx}}
		\left(
		(\widehat{\mathbf{F}}_{M_R}
		(\boldsymbol{\theta}_j^p))_{j=1}^{N_p}
		\right)
		\right\}.
	\end{equation}
	The whole process is summarized in the following Algorithm \ref{alg:qehvi_edgecase}.
	\begin{algorithm}[t]
		\caption{qEHVI-Based Discovery of Estimator-Stealthy Adversarial Edge Cases}
		\label{alg:qehvi_edgecase}
		\begin{algorithmic}[1]
			\Require Frozen victim policy $\pi_{\theta_v}$; attack parameter domain \(\Theta=[-\theta_{\max},\theta_{\max}]^{\tilde{d}}\) ;  NIS threshold $\tau_{D}$; initial design size $N_{0}$; BO iterations $T^{\mathrm{BO}}$;
			batch size $q$; search budget $B$; rollout budget $M$; Pareto re-evaluation budget $M_{\mathrm{R}}$
			\Ensure Confirmed empirical Pareto set \(\Theta_{\mathrm{conf}}^{\star}\).
			\State Evaluate the unattacked system over \(M\) independent closed-loop episodes and estimate the nominal expected return $\hat J_{\mathrm{clean}} = \frac{1}{M} \sum_{m=1}^{M} J_{\mathrm{clean}}^{(m)}.$
			\State Generate an initial space-filling design $ \Theta_0 = \left\{ \boldsymbol{\theta}_1,\ldots,\boldsymbol{\theta}_{N_0} \right\} \subset\Theta$.
			\State For each \(\boldsymbol{\theta}_n\in\Theta_0\), execute \(M\) independent closed-loop episodes with attacked GNSS measurements as per (\ref{eq:43}).
			\State Estimate the navigation-impact objective as (\ref{eq:45}), while considering the average return over $M$ independent episodes.
			\State For each valid GNSS correction \(i\), compute the NIS threshold-violation indicator as per (\ref{eq:46}) to formulate the detectability objective.
			\State Formulate the corresponding multi-objective function as per (\ref{eq:48}), while constructing an initial observation set $\mathcal D_0 = \left\{ (\mathbf x_n,\mathbf y_n) \right\}_{n=1}^{N_0}, $ where $\mathbf{x}_n$ is defined as per (\ref{eq:51}).
			\While{the prescribed search budget $B$ is not exhausted}
				
				\State Fit or update the probabilistic surrogate using all available observations as per (\ref{eq:52}).
				\State Determine the current non-dominated objective set \(\mathcal P_t\).
				\State Select the next batch of \(q\) attack candidates by maximizing the qEHVI acquisition function, as per (\ref{eq:54}) and (\ref{eq:55}).
				\State Map each normalized candidate \(\mathbf x^{(j)}\in\mathbf X_q^{\star}\) back to the physical attack-parameter space, 
				\[\boldsymbol{\theta}^{(j)} = \theta_{\max} \left( 2\mathbf x^{(j)}-\mathbf 1 \right). \]
				\State Evaluate each \(\boldsymbol{\theta}^{(j)}\) over \(M\) independent closed-loop episodes and compute the empirical estimates as per (\ref{eq:56}).
				\State Augment the observation set as  $\mathcal D_{t+1} = \mathcal D_t \cup \left\{ \left( \mathbf x^{(j)},
				\hat{\mathbf F}(\boldsymbol{\theta}^{(j)}) \right) \right\}_{j=1}^{q}.$				
			\EndWhile 
			\State Extract the provisional non-dominated attack set from all evaluated candidates  as per (\ref{eq:57}) and re-evaluate every \(\boldsymbol{\theta}\in\widehat{\Theta}^{\star}\) using the larger independent rollout budget \(M_R\).
			\State Perform non-dominated sorting on the re-evaluated objective values and obtain $\Theta_{\mathrm{conf}}^{\star}$ as per (\ref{eq:57}).
			\State \Return $\Theta_{\mathrm{conf}}^{\star}$.	
		\end{algorithmic}
	\end{algorithm}
	\section{Results and Discussions}
	\subsection{Numerical Experiments}
	The numerical experiments are conducted using a GPU-accelerated, vectorized quadrotor simulation framework \cite{dp457_sample_efficient_ddpg_quadrotor}. The navigation policy is trained using demonstration trajectories generated by a classical goal-seeking and obstacle-avoidance policy \cite{nair2018overcoming} to improve training efficiency and convergence. These demonstrations are used to pre-train the actor network through behavioural cloning, reducing the effective number of samples for exploration during subsequent policy updates. The hyperparameters to recreate the RL training and convergence results are provided in our GitHub repository \cite{dp457_sample_efficient_ddpg_quadrotor}. 
	\subsection{DDPG Training Convergence and Nominal Navigation Performance}
	\begin{figure}[!t]
		\centering
		\includegraphics[width=0.50\textwidth]{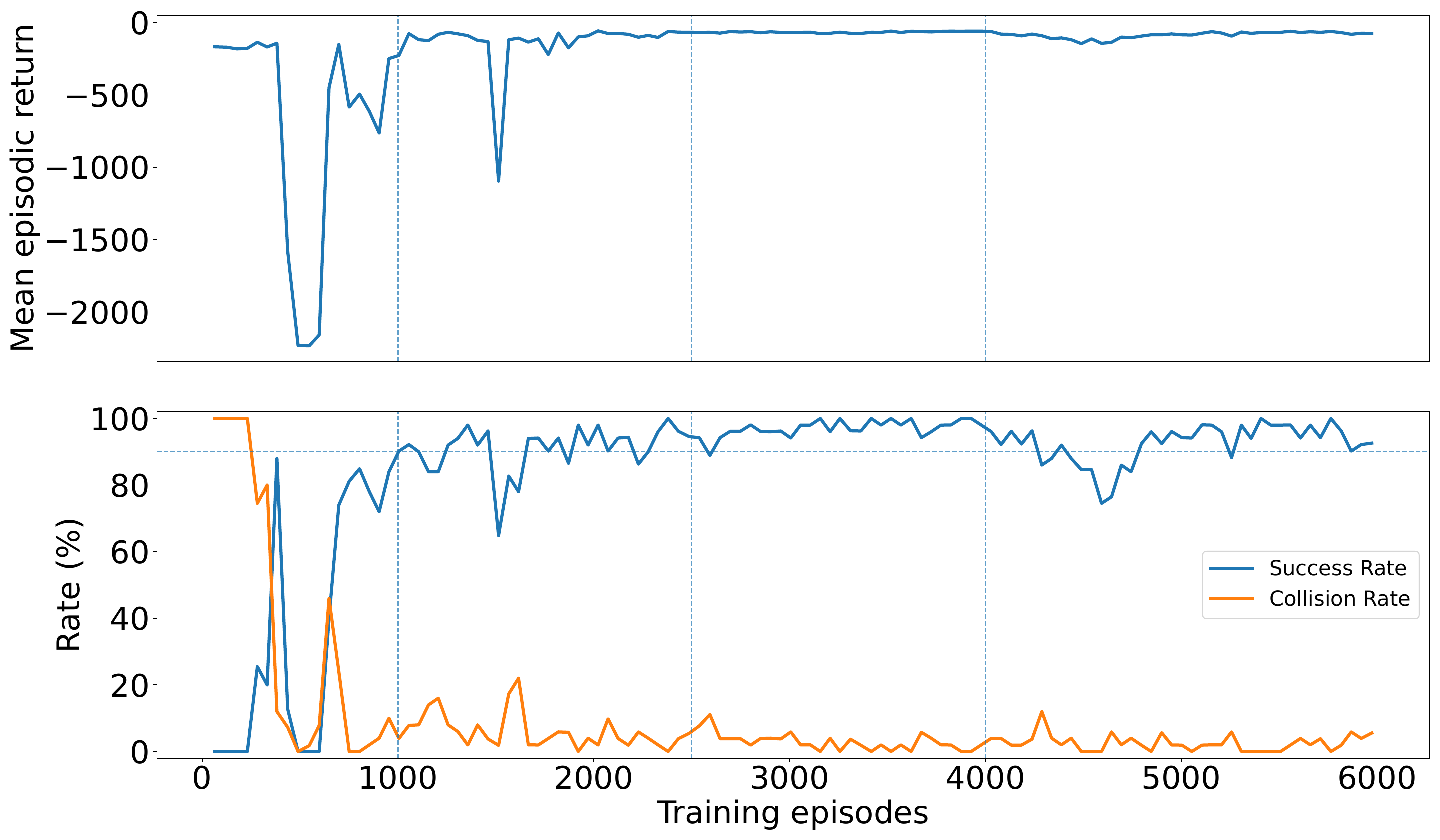}
		\caption{ Training performance of the DDPG-based UAV navigation policy over 6000 episodes. The upper panel shows the mean episodic return, while the lower panel presents the corresponding mission success and collision rates.}
		\label{figure2_convergence}
	\end{figure}
	As illustrated in Fig. \ref{figure2_convergence}, the training performance of the DDPG-based navigation policy is observed over 6000 episodes while considering its episodic reward and mission success rate. During the initial training phase, episodic returns fluctuate considerably, accompanied by frequent collisions and mission failures. As training progresses, the mean episodic return gradually stabilizes, with mission success rates exceeding 90\% and collision rates remaining below 5\% during the later stages. These trends demonstrate stable goal-directed navigation under nominal conditions, establishing a performance baseline for assessing GNSS spoofing-induced navigation degradation.
	\subsection{Estimator Innovation Residuals and Detectability Bounds}
	\begin{figure}[!t]
		\centering
		\includegraphics[width=0.50\textwidth]{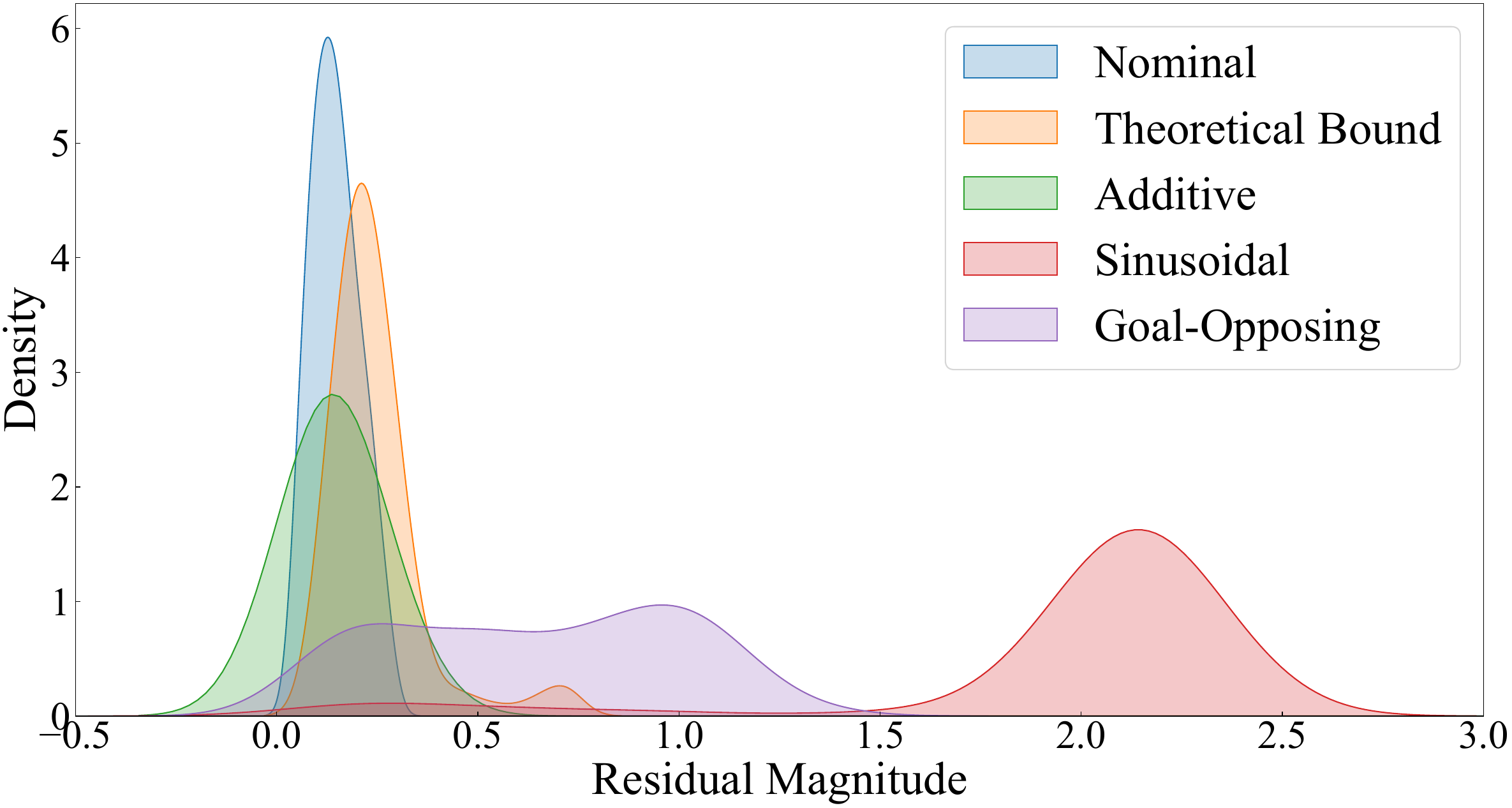}
		\caption{Kernel density estimates of innovation-residual magnitudes under nominal conditions and different perturbation strategies, including additive, sinusoidal, and goal-opposing attacks, alongside the empirical distribution of theoretical-bounds evaluated at successive GNSS correction instants.}
		\label{figure4_distributional_bound}
	\end{figure}
	The theoretical innovation-residual bounds are evaluated at successive GNSS correction instants, yielding a sequence of deterministic bounds. As illustrated in Figure \ref{figure4_distributional_bound}, the residual distributions exhibit distinct characteristics under nominal and adversarial conditions, with the former sharply concentrated at low residual magnitudes, overlapping with the distribution induced from the theoretical-bound. The rightward shift in the residual distribution under sinusoidal perturbations indicates increased inconsistency between GNSS measurements and estimator predictions, whereas the broader distribution shift associated with goal-opposing perturbations suggests a more heterogeneous estimator response. In contrast, the innovation residual distribution under additive perturbations exhibits significant overlap with the reference distribution obtained from the theoretical bounds. These observations suggest that estimator-level statistical consistency depends not only on perturbation magnitude but also on perturbation structure, motivating the subsequent multi-objective assessment of estimator detectability and downstream navigation robustness.
	\begin{figure}[!t]
		\centering
		\includegraphics[width=0.50\textwidth]{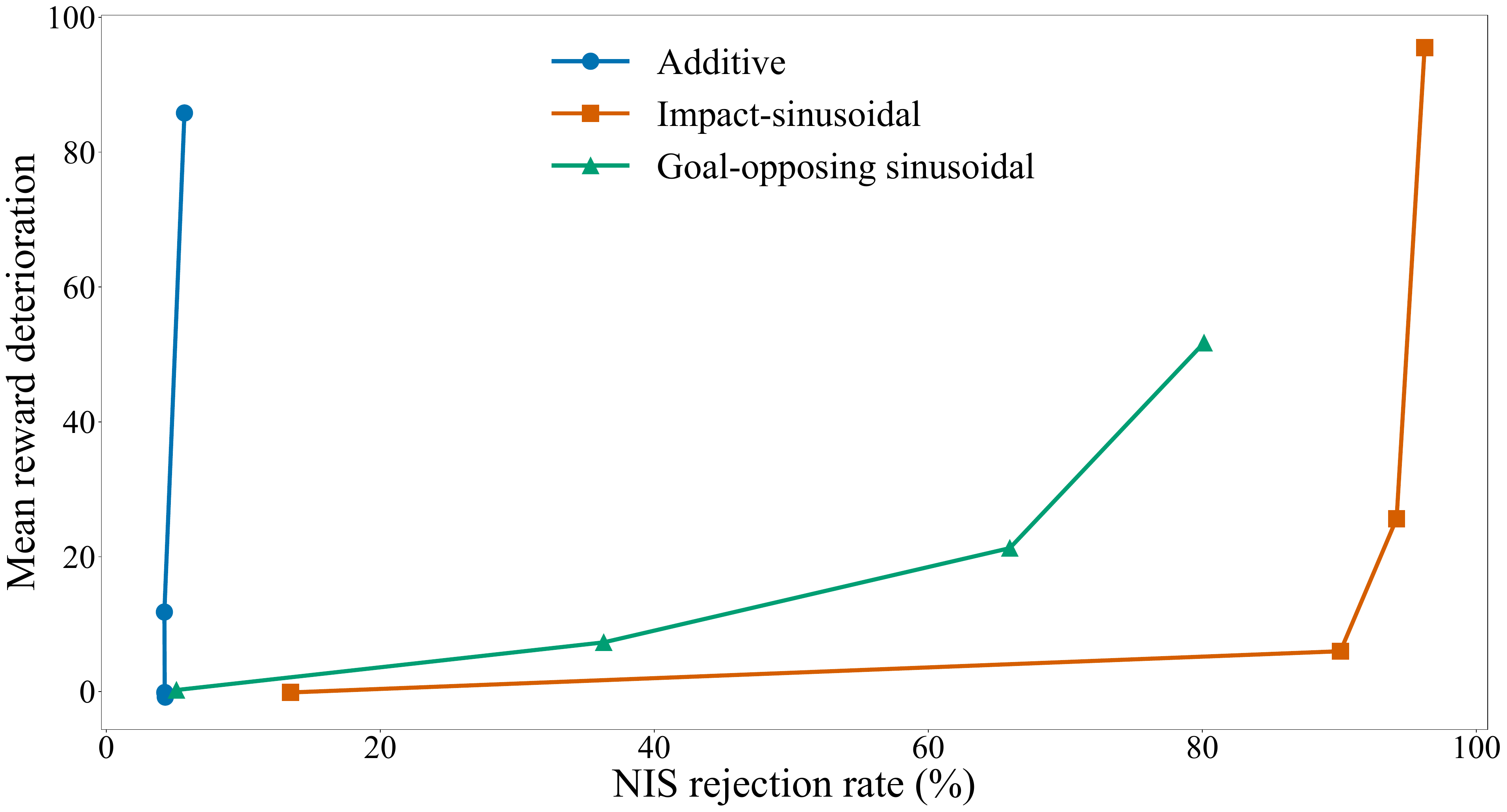}
		\caption{Relationship between expected reward deterioration and NIS rejection rate under additive, impact-sinusoidal, and goal-opposing sinusoidal GNSS spoofing attacks, with increasing attack impact.}
		\label{figure5_reward_detectability}
	\end{figure}
		
	As shown in Figure \ref{figure5_reward_detectability}, we observe an attack-dependent relationship between mission-level degradation and estimator-level detectability where larger reward deterioration, as per (\ref{eq:46}), is generally associated with increased NIS rejection rates, as per (\ref{eq:45}), for the impact-sinusoidal and goal-opposing sinusoidal attacks, the additive perturbation exhibits a noticeably different response, as it induces significant reward deterioration at low rejection rates. These empirical findings support the local policy-sensitivity analysis in Theorem \ref{thm2}, linking estimator-admissible perturbations to expected reward degradation. The next subsection examines this cross-layer propagation through the sensitivity matrix defined in (\ref{eq:37}).
	\subsection{Cross-Layer Policy Sensitivity to Perturbations}
		\begin{figure}[!t]
		\centering
		\includegraphics[width=0.50\textwidth]{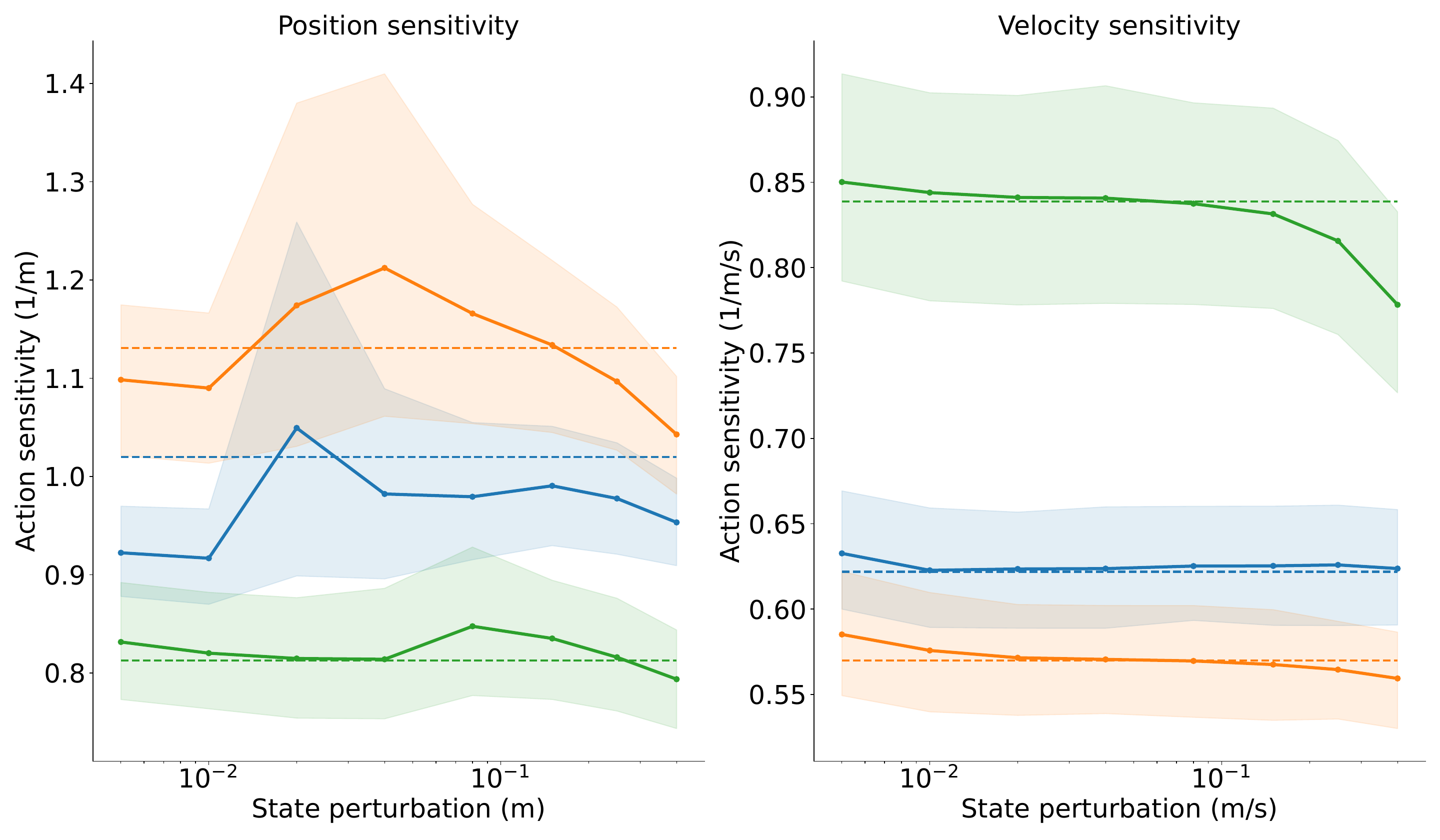}
		\caption{Empirical sensitivity of the trained DDPG navigation policy to axis-specific perturbations in EKF-estimated position (left) and velocity (right). Solid curves represent the mean action deviation normalized by perturbation magnitude, dashed lines indicate local Jacobian-based sensitivity estimates obtained using central finite differences, and shaded regions denote 95\% trajectory-cluster bootstrap confidence intervals.}
		\label{figure5_sensitivity}
		\end{figure}						
	The sensitivity of the learned navigation policy to perturbations in the EKF-estimated states is investigated to empirically characterise the estimator-to-policy sensitivity underlying the cross-layer formulation in (\ref{eq:37}). The analysis employs a trained DDPG actor with fixed parameters across 64 parallel navigation environments, each containing three obstacles. Position (0.005-0.4 m) and velocity (0.005–0.4 m/s) perturbations are independently applied along each spatial axis while keeping the physical UAV state and estimator covariance unchanged. Empirical action sensitivity is quantified as the Euclidean norm of the resulting action deviation normalised by the perturbation magnitude, with local reference sensitivities estimated using central finite differences with a step size of 0.02 in the corresponding physical units. Sensitivity estimates are averaged across sampled navigation states, with 95\% confidence intervals obtained through trajectory-cluster bootstrap resampling as shown in Figure \ref{figure5_sensitivity}.  he results characterise the navigation policy’s response to directional perturbations, with empirical sensitivities remaining relatively consistent with the local references at smaller perturbation magnitudes.
	\begin{figure}[!t]
		\centering
		\includegraphics[width=0.50\textwidth]{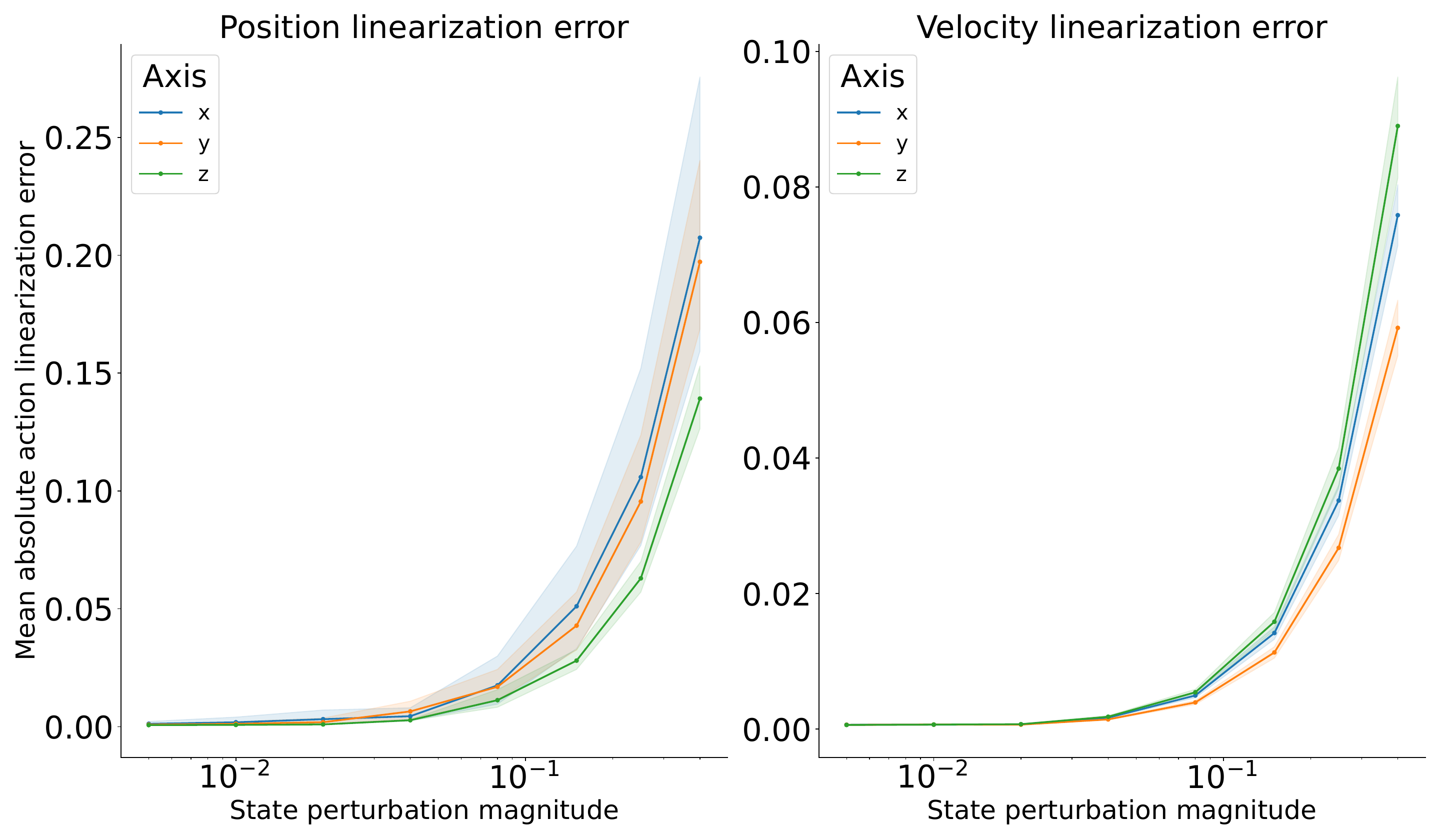}
		\caption{Mean absolute action linearization error under axis-specific perturbations of the EKF-estimated position (left) and velocity (right). The error is computed as the Euclidean norm of the difference between actual action deviations and their first-order Jacobian-based predictions.}
		\label{figure6_linearization_error}
	\end{figure}
	 
	 The validity of the first-order approximation in (\ref{eq:37}) is further assessed by quantifying the Euclidean norm of the difference between the observed and predicted action deviations, as shown in Figure \ref{figure6_linearization_error}. We observe that the linearisation error remains small for low-magnitude perturbations but increases nonlinearly as their intensity grows. These findings empirically support the first-order approximation employed in Theorem \ref{thm2}, where the cross-layer sensitivity matrix $\mathbf{M}_i$ characterises the propagation of measurement perturbations. However, increasing nonlinear effects and potential violations of estimator-admissibility conditions at larger perturbation magnitudes restrict the validity of the theoretical characterisation of the impact-detectability trade-off. This motivates the subsequent closed-loop multi-objective optimisation framework, focusing on identifying low-magnitude adversarial perturbations that induce reward degradation while preserving estimator-level admissibility.
	 
	\subsection{Multi-Objective Optimization for Adversarial Edge-Case Identification}
	In order to solve the multi-objective optimization framed in (\ref{eq:48}), three search strategies are considered in this paper: multi-objective Bayesian optimization based on qEHVI \cite{daulton2021parallel}, non-dominated sorting genetic algorithm II (NSGA-II) \cite{deb2002fast}, and multi-dimensional archive of phenotypic elites (MAP-Elites) \cite{pierrot2022multi}. All three methods optimize the parameters of the same bounded, telemetry-dependent GNSS perturbation policy, while the victim DDPG navigation policy remains fixed. This comparison examines both the efficiency of discovering consequential attacks under estimator-level detectability limits and the quality of the resulting impact–detectability trade-off for different search strategies.
	
\begin{figure}[!t]
		\centering
		\includegraphics[width=0.50\textwidth]{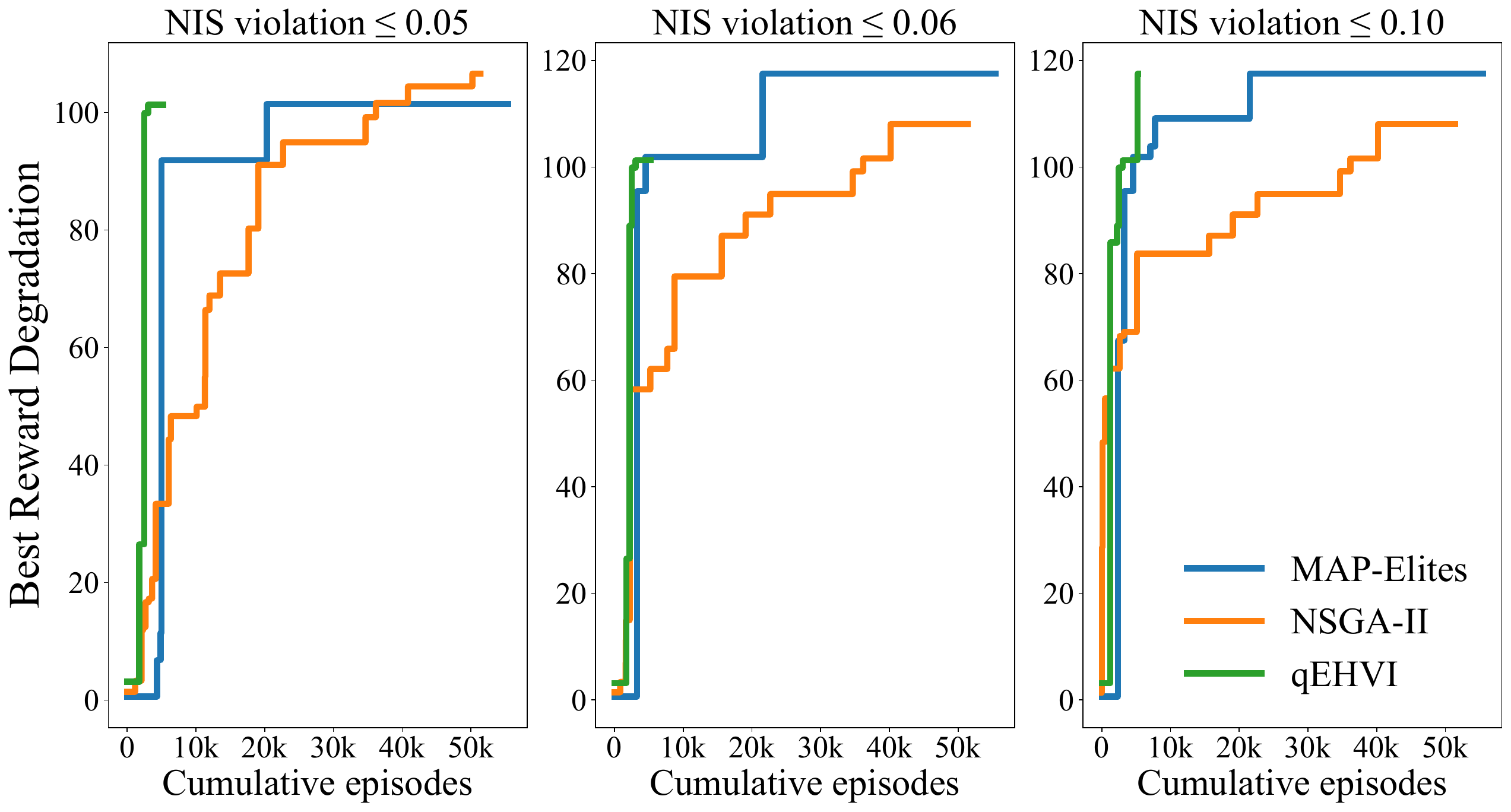}
		\caption{Best return degradation achieved by MAP-Elites, NSGA-II, and qEHVI as a function of cumulative evaluated episodes under maximum NIS violation rates of (a) 0.05, (b) 0.06, and (c) 0.10.}
		\label{figure7_reward_degrade}
	\end{figure}

\begin{figure}[!t]
		\centering
		\includegraphics[width=0.50\textwidth]{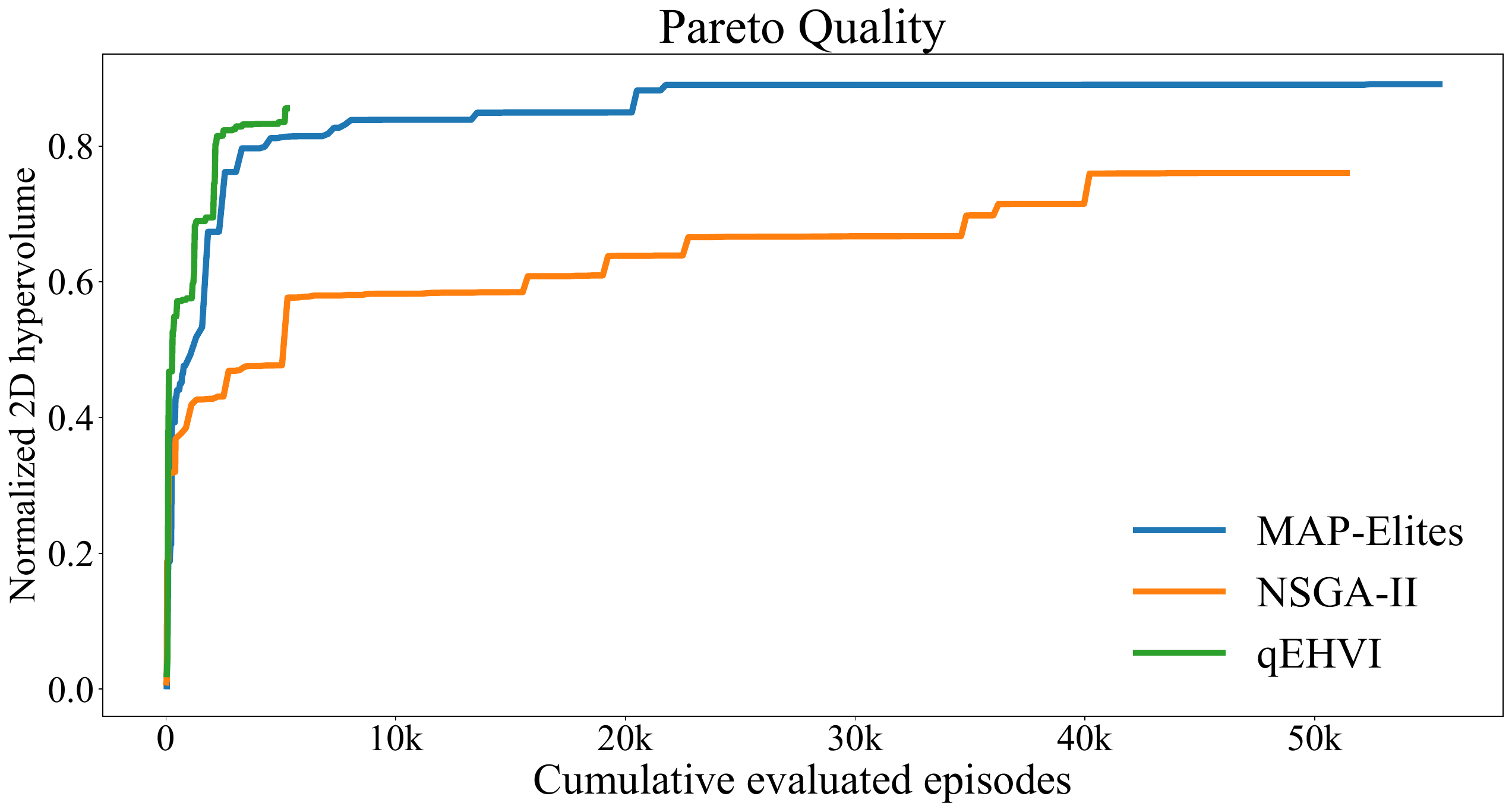}
		\caption{Evolution of normalized two-dimensional hypervolume for MAP-Elites, NSGA-II, and qEHVI against cumulative evaluated episodes.}
		\label{figure8_pareto_quality}
	\end{figure}

	As shown in Figure \ref{figure7_reward_degrade}, maximum observed reward degradation attained over cumulative evaluation episodes is compared against the maximum NIS violation rates of 0.05, 0.06, and 0.10. Across all thresholds, MAP-Elites and qEHVI identify high-impact perturbations earlier than NSGA-II, with qEHVI achieving comparable reward degradation within a smaller evaluation budget. Relaxing the violation threshold enables the discovery of additional high-impact perturbations without altering the relative performance of the three methods. While these results characterise search efficiency under prescribed detectability constraints, they do not capture the quality of the discovered solutions across the joint impact-detectability objective space. This is assessed using the normalised two-dimensional hypervolume in Fig. \ref{figure8_pareto_quality}, where MAP-Elites and qEHVI exhibit faster convergence than NSGA-II. MAP-Elites achieves the highest final hypervolume of approximately 0.9, followed closely by qEHVI, whereas NSGA-II reaches approximately 0.75.
	
\begin{figure}[!t]
		\centering
		\includegraphics[width=0.50\textwidth]{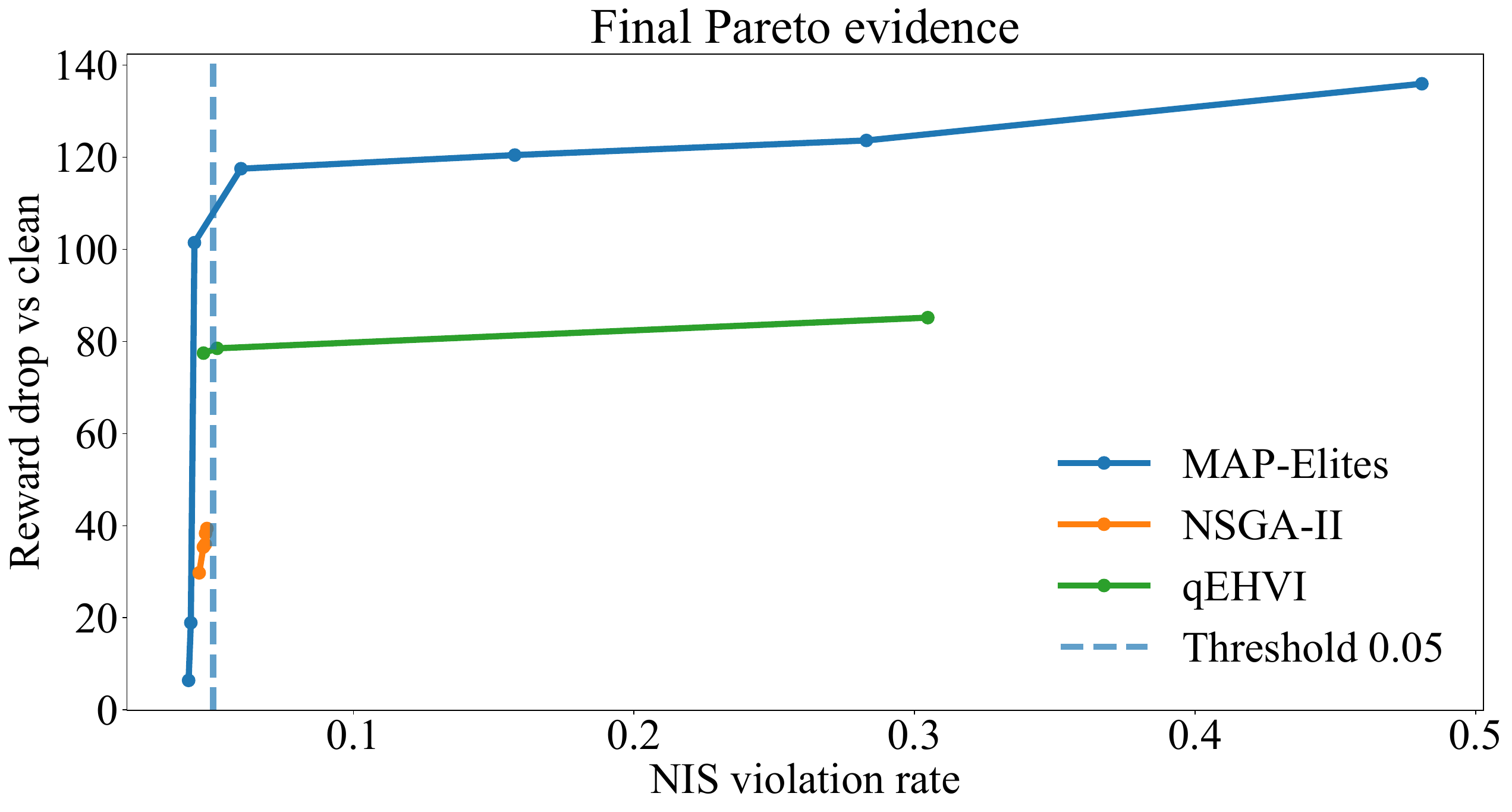}
		\caption{Final reward degradation relative to nominal navigation performance versus NIS violation rate for adversarial candidates identified by MAP-Elites, NSGA-II, and qEHVI. The dashed vertical line denotes the nominal NIS violation threshold of 0.05.}
		\label{figure9_pareto_front}
	\end{figure}

	The distribution of solutions in Fig. \ref{figure9_pareto_front} provides further insight into the observed hypervolume differences. Although MAP-Elites achieves a marginally higher hypervolume, its solutions span a wider range of NIS violation rates, with the greatest reward degradation associated with higher detectability. In contrast, qEHVI identifies solutions concentrated within a low-detectability region, achieving substantial reward degradation in expected nominal mission reward while maintaining NIS violation rates close to the prescribed threshold of 0.05. This distinction reflects the underlying search mechanisms: MAP-Elites promotes behavioural diversity, whereas qEHVI directly optimises the expected improvement in the joint objective space. Consequently, despite its slightly lower hypervolume, qEHVI provides a more targeted set of high-impact, low-detectability perturbations. These solutions are therefore selected for subsequent evaluation of their detectability against a broader class of detection algorithms, while evaluating the consequences for mission-level navigation performance.
	\subsection{Cross-Layer Detectability and Mission-Level Consequences of Adversarial Edge Cases}
	In order to assess whether the identified adversarial edge cases remain difficult to detect beyond the criterion considered during multi-objective  optimization, as described in the previous subsection, the Pareto-optimized GNSS perturbations are benchmarked against conventional additive, impact-oriented sinusoidal, and goal-opposing sinusoidal attacks. In this subsection, we consider complementary monitoring mechanisms to detect the attacks
	that considers the estimator-residual and navigation trajectory levels as the input. At the estimator level, cumulative sum (CUSUM) \cite{basseville1993detection}, statistical distribution-based screening (SDS) \cite{basseville1993detection}, chi-squared NIS testing \cite{bar2001estimation}, and GRU-based temporal variational autoencoder (GRU-TVAE) \cite{park2018multimodal, wang2025reliability} are evaluated using estimator innovation residual. At the navigation-state level, GRU autoencoder (GRU-AE) \cite{malhotra2016lstm, zhan2025seq2seq}, GRU-TVAE \cite{park2018multimodal}, and one-class support vector machines (OC-SVM) \cite{ji2022event} are employed to characterize deviations from nominal navigation behaviour. 
	\begin{figure}[!t]
		\centering
		\includegraphics[width=0.50\textwidth]{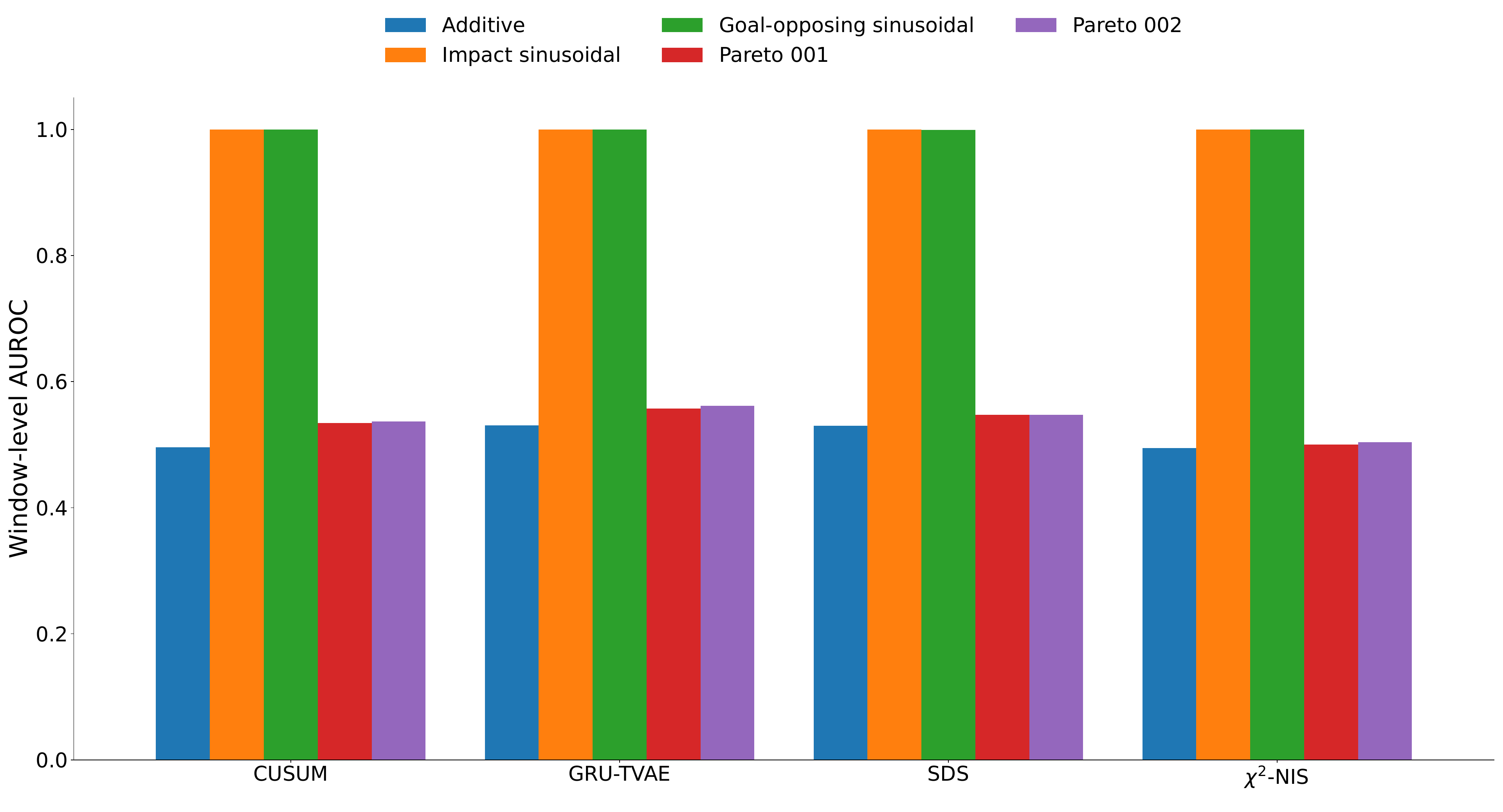}
		\caption{Window-level detection AUROC for conventional GNSS spoofing attacks and two Pareto-optimized adversarial policies across CUSUM, GRU-TVAE, SDS-inspired statistical screening, and chi-squared NIS detectors. Values approaching 0.5 indicate near-random discrimination between nominal and attacked innovation sequences.}
		\label{figure10_AURIC_residual}
	\end{figure}		
	
	As shown in Figure \ref{figure10_AURIC_residual}, conventional impact-oriented and goal-opposing sinusoidal attacks
	exhibit can be detected almost perfectly, with AUROC values approaching 1.0 across the considered detectors. In contrast, both Pareto-optimized attacks yield relatively lower AUC-ROC values, approximately 0.50-0.56, indicating limited detectability between nominal and attacked innovation sequences. This behaviour is also observed consistently across CUSUM, GRU-TVAE, SDS, and chi-squared based detectors, which exhibit different statistical assumptions behind the anomaly-scoring mechanisms. The additive attack also exhibits relatively low detectability, but its downstream consequences on the mission are less severe than those induced by the Pareto attacks. 
	\begin{figure}[!t]
		\centering
		\includegraphics[width=0.50\textwidth]{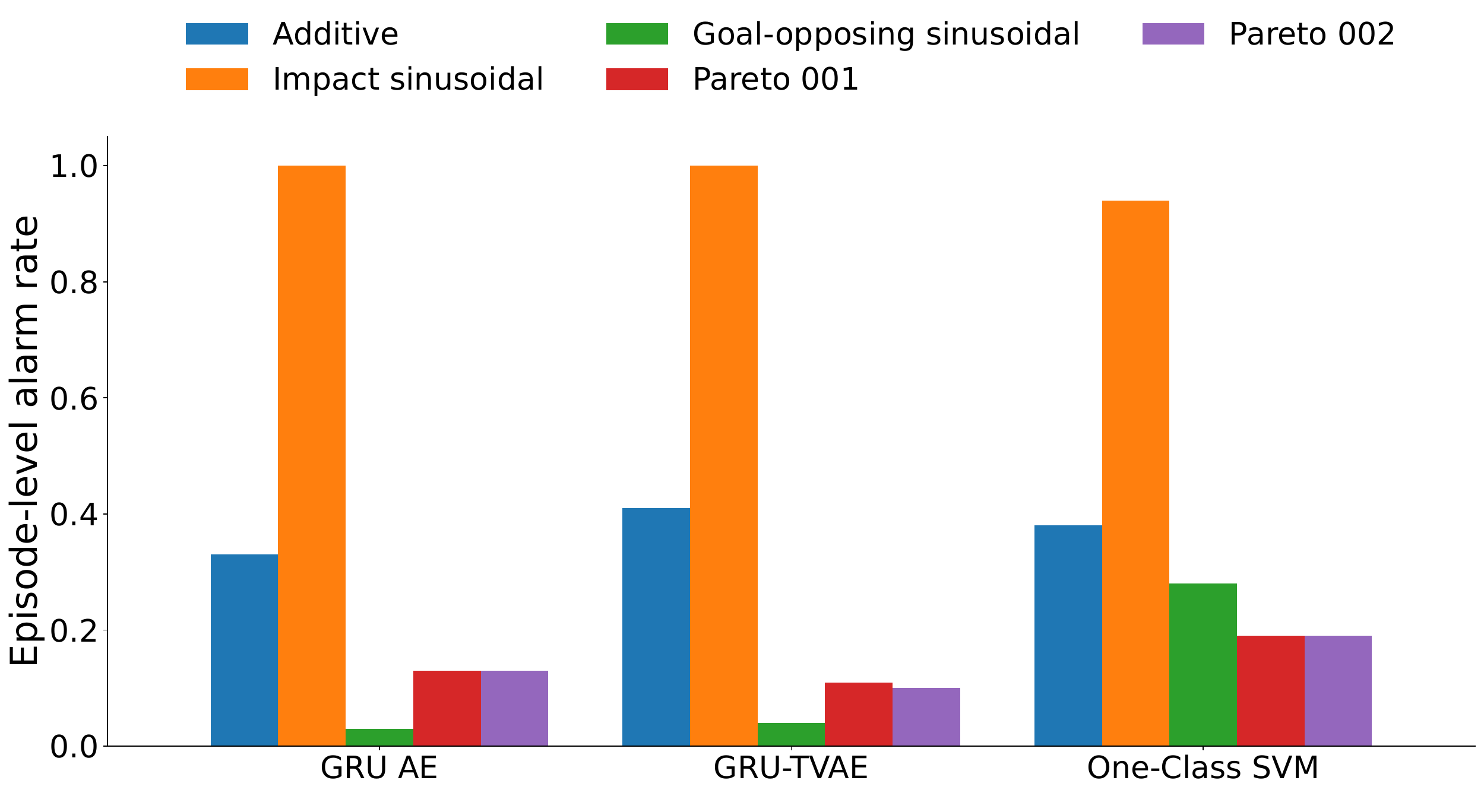}
		\caption{Episode-level alarm rates for additive, sinusoidal, goal-opposing, and Pareto-optimized GNSS spoofing attacks using GRU-AE, GRU-TVAE, and one-class SVM detectors.}
		\label{figure11_state_ad}
	\end{figure}
	
	As per the episode-level alarm rates obtained from navigation-state anomaly detection mechanisms shown in Figure \ref{figure11_state_ad}, the detection rate of the impact-oriented sinusoidal attack is nearly perfect in all evaluated episodes by the recurrent autoencoder models, while the additive attack produces alarm rates of approximately 30-40\%. In contrast, the Pareto-optimized attacks achieve substantially lower alarm rates of approximately 10-15\% for GRU-AE and GRU-TVAE, despite inducing considerable mission-level degradation. The one-class SVM exhibits a similar trend, with alarm rates of approximately 20\% for Pareto-optimized attacks, compared to the higher detection rates for the impact-oriented sinusoidal attack. 
	
	Although additive and Pareto-optimized attacks of comparable magnitudes exhibit similar detectability under certain conditions, the Pareto-optimized attacks are less frequently detected in the RL navigation state-space while producing severe downstream consequences. 
	\begin{figure}[!t]
		\centering
		\includegraphics[width=0.50\textwidth]{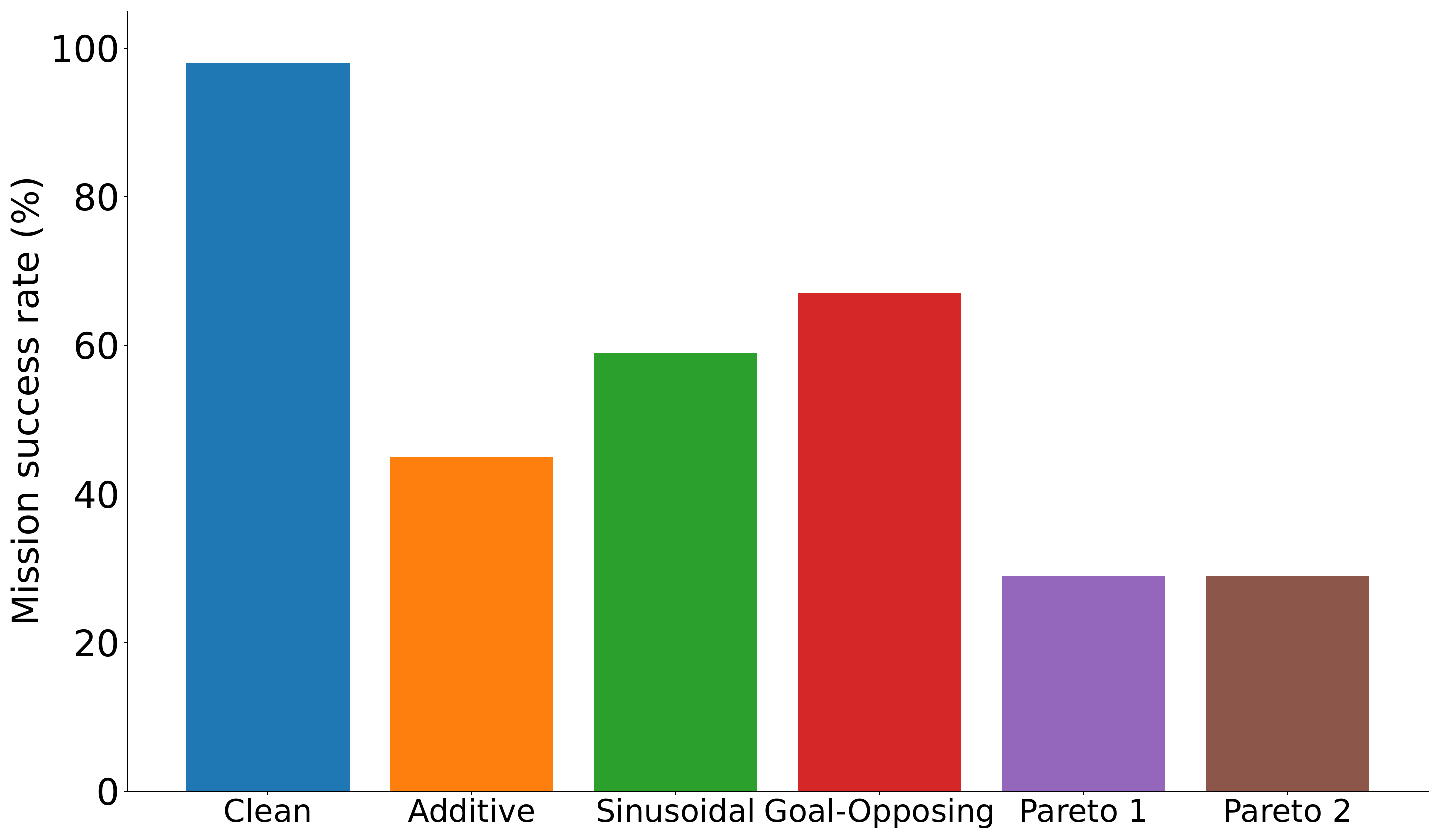}
		\caption{Mission success rates under nominal operation and conventional and Pareto-optimized GNSS spoofing attacks.}
		\label{figure12_mission_success}
	\end{figure}					
	\begin{figure}[!t]
		\centering
		\includegraphics[width=0.50\textwidth]{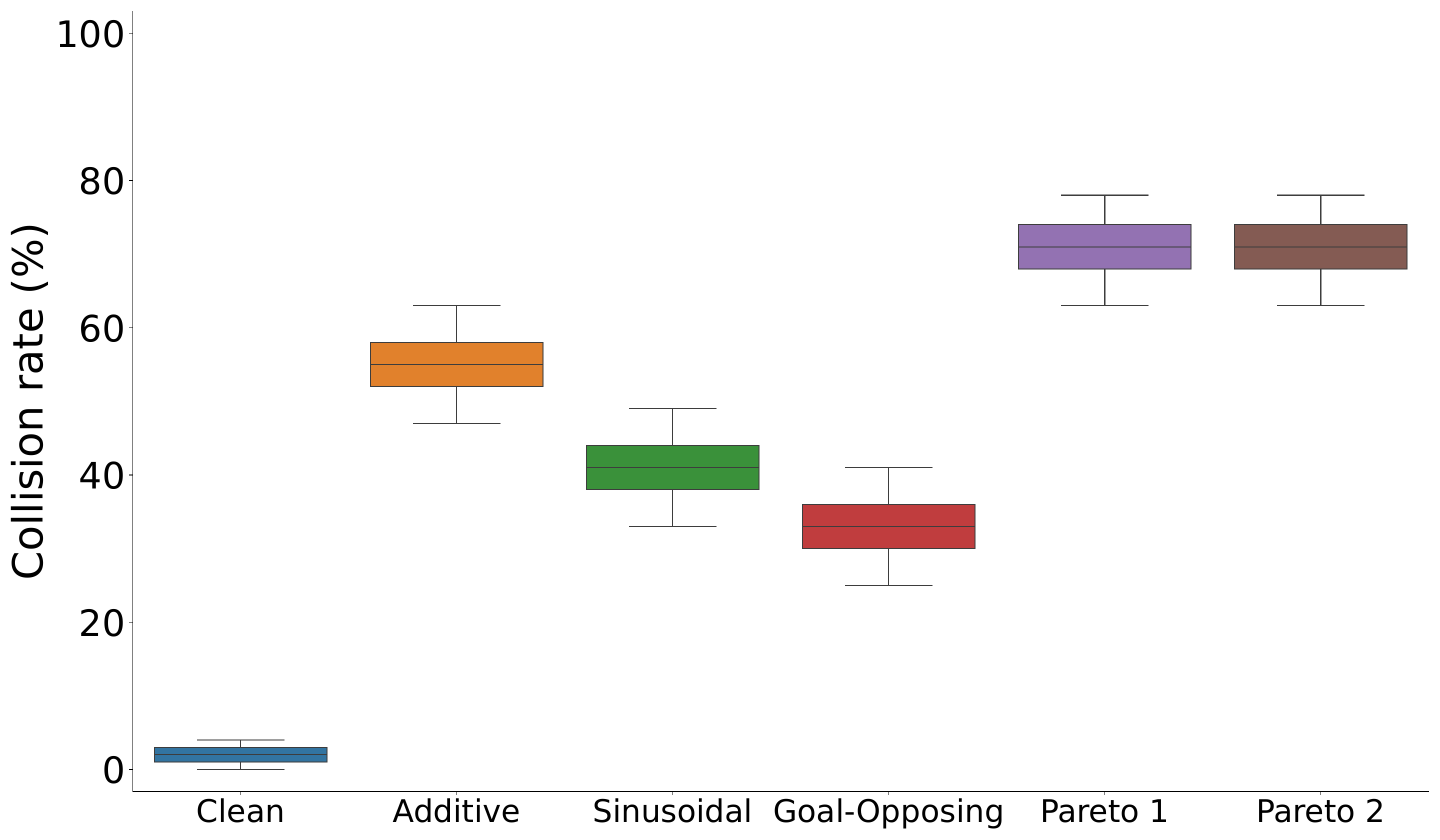}
		\caption{Distribution of collision rates across evaluation batches under nominal operation and under different GNSS spoofing attacks.}
		\label{figure13_collision_rate}
	\end{figure}
	Mission-level performance is evaluated across 100 evaluation batches, where the collision rate for each batch is computed as the proportion of environments experiencing a collision, while mission success is quantified as the proportion of environments successfully reaching the navigation goal. Figure \ref{figure12_mission_success} presents the aggregate mission success rates, whereas Fig. \ref{figure13_collision_rate} characterises the distribution of collision rates across evaluation batches. Under nominal operation, the UAV achieves a mission success rate of approximately 98\%, with a median collision rate close to 2\%. Conventional additive, sinusoidal, and goal-opposing attacks reduce mission success to approximately 45\%, 59\%, and 67\%, respectively. In contrast, both Pareto-optimized attacks reduce mission success to approximately 28-29\%, representing the most severe degradation among the evaluated conditions. A corresponding increase in collision risk is observed in Fig. \ref{figure13_collision_rate}, where the Pareto attacks produce median collision rates of approximately 70-72\%, exceeding those observed under conventional additive, sinusoidal, and goal-opposing attacks, respectively.  
		\begin{figure}[!t]
		\centering
		\includegraphics[width=0.50\textwidth]{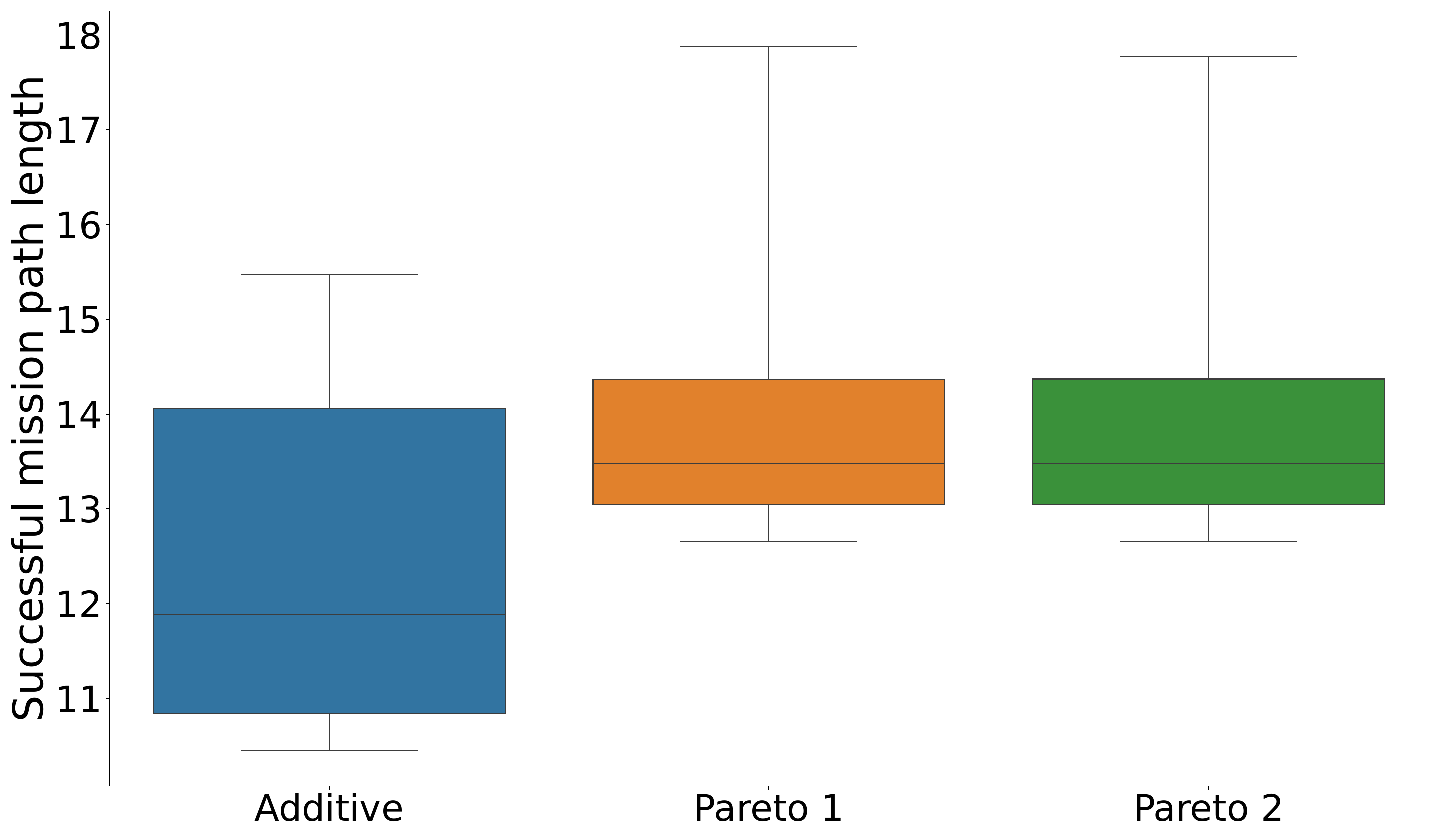}
		\caption{Distribution of successful mission path lengths under additive and Pareto-optimized GNSS spoofing attacks, illustrating the effects of adversarial perturbations on navigation efficiency conditional on successful mission completion.}
		\label{figure14_path_length}
	\end{figure}
	If we compare the path lengths of successful missions of additive perturbations with the Pareto-optimized attacks in Figure \ref{figure14_path_length} across 100 evaluated episodes, the results indicate that the latter yield higher median path lengths than the former, suggesting reduced path efficiency even when the UAV successfully reaches its destination. Although the increase in median path length is modest, this finding highlights an additional consequence of Pareto-optimized attacks. Beyond their ability to reduce detectability, increase collision rates, and compromise mission success, these attacks also degrade the efficiency of successfully completed missions by inducing longer trajectories. 
	
	Collectively, Figs. \ref{figure10_AURIC_residual}-\ref{figure14_path_length} demonstrate that estimator-level statistical consistency and state-space anomaly monitoring alone cannot guarantee downstream navigation integrity. Unlike conventional sinusoidal perturbations, Pareto-optimised attacks exploit the divergence between estimator-level detectability and policy sensitivity, inducing substantial mission degradation while eliciting weaker detector responses. These observations are consistent with Theorem \ref{thm2}, which establishes conditions linking estimator-admissible perturbations to downstream policy degradation. Furthermore, the model-independent detectability formulation established in Theorem \ref{th:1} provides a theoretical basis for investigating such vulnerabilities across different UAV platforms and open-source flight control systems. These findings motivate cross-layer monitoring frameworks that explicitly account for the propagation of estimation errors into navigation decisions and mission-level outcomes.
	\subsection{TD-Space Characterization of Edge Cases}
\begin{figure}[!t]
		\centering
		\includegraphics[width=0.50\textwidth]{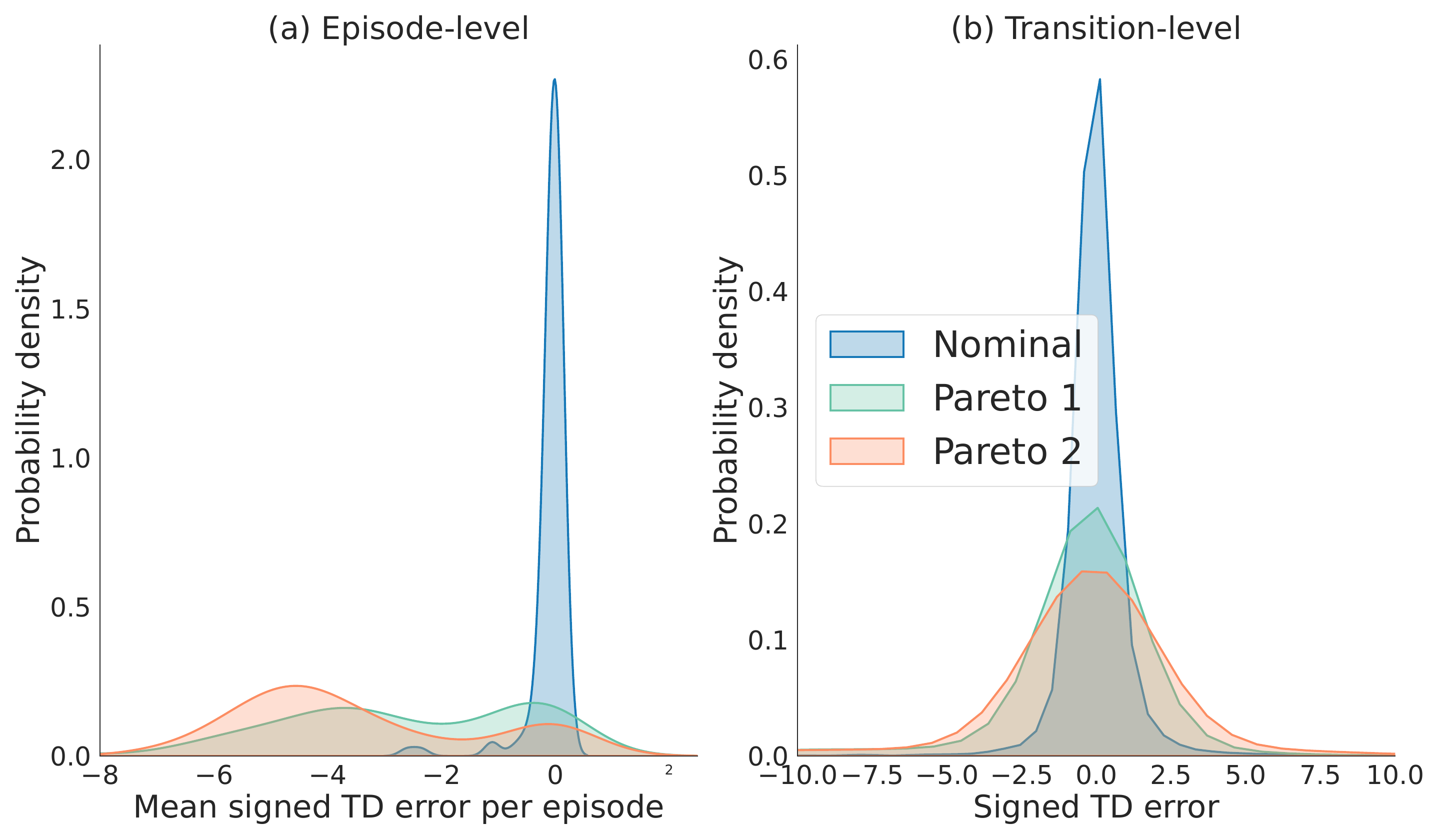}
		\caption{Signed temporal-difference (TD) error distributions under nominal navigation and two Pareto-optimised GNSS spoofing attacks: (a) episode-level mean signed TD error and (b) transition-level signed TD error. The negative distributional shifts at the episode level suggest the potential utility of downstream value-function consistency as a complementary monitoring indicator.}
		\label{figure15_td_error_detection}
	\end{figure}

	While Theorem \ref{thm2} establishes sufficient local conditions under which estimator-admissible perturbations induce expected mission-return degradation, we further investigate whether such degradation manifests in downstream value-function consistency. Motivated by our previous work \cite{panda2026curr}, we consider the signed temporal-difference (TD) error of the trained DDPG agent, defined as
	\[
	\begin{aligned}
		\delta_k^{\mathrm{TD}} = {}&r_k+\gamma(1-d_k)Q_{\bar{\psi}}\left( \boldsymbol{\Phi}_{k+1}, \pi_{\bar{\theta}}(\boldsymbol{\Phi}_{k+1}) \right)\\
		&-Q_\psi(\boldsymbol{\Phi}_k,\mathbf a_k),
	\end{aligned}
	\]
	where \(Q_\psi\) and \(Q_{\bar{\psi}}\) denote the trained critic and target critic, respectively, \(\pi_{\bar{\theta}}\) is the target actor, and \(d_k\) denotes the terminal-transition indicator. The network parameters remain fixed throughout the evaluation. The episode-level value-prediction consistency is characterised using the mean signed TD error,
	\[
		\overline{\delta}^{\mathrm{TD}} = \frac{1}{T}\sum_{k=0}^{T-1} \delta_k^{\mathrm{TD}},
	\]
	where \(T\) denotes the episode length. As illustrated in Fig. \ref{figure15_td_error_detection}(a), nominal navigation exhibits a sharply concentrated episode-averaged TD-error distribution near zero, whereas both Pareto-optimised attacks induce broader distributions shifted towards negative values, with Pareto 2 exhibiting a more pronounced negative tail. This indicates a systematic tendency towards lower realised Bellman targets relative to the critic's predictions. In contrast, Fig. \ref{figure15_td_error_detection}(b) reveals considerable overlap between nominal and adversarial TD-error distributions at the transition level. These findings suggest that temporal aggregation of signed TD errors exposes deviations in value-prediction consistency that are less distinguishable at individual transitions.
	
	The observed distributional shifts complement the limited detectability of the Pareto-optimised attacks at the estimation and state-space monitoring layers, motivating the investigation of downstream value-consistency indicators. Such indicators could reduce reliance on labelled adversarial samples, particularly in operational settings where the rarity of adversarial edge cases introduces class-imbalance challenges for supervised classification \cite{panda2023hyperparameter}. However, the reward function in (\ref{eq:8}) incorporates true geometric clearance, which may not be directly available onboard. Therefore, these findings demonstrate the potential utility of TD-space monitoring rather than establish a deployable detection mechanism.
	\section{Conclusion}
	This paper investigated the cross-layer vulnerability of learning-enabled UAV navigation to adversarial GNSS spoofing, focusing on the relationship between estimator-level statistical detectability and downstream mission integrity. A plant-model-independent analytical framework was developed to characterize multi-rate estimation-error propagation and innovation-residual bounds using the estimator prediction–correction structure without requiring complete knowledge of the nonlinear UAV dynamics. Furthermore, sufficient local conditions were established under which estimator-admissible perturbations can induce downstream navigation-performance degradation through the sensitivity of the learned policy. Building on these theoretical results, a multi-objective Bayesian optimization framework was formulated to systematically discover Pareto-efficient adversarial edge cases that maximize mission degradation while minimizing estimator-level detectability. The proposed approach was evaluated using a DDPG-based quadrotor navigation system operating in dynamic airspace environments, with comparisons against conventional additive, sinusoidal, and goal-opposing GNSS spoofing attacks.
	
	The simulation results demonstrated that Pareto-optimized attacks reduced mission success from approximately 98\% under nominal conditions to 28–29\%, while increasing median collision rates from approximately 2\% to 70–72\%. Despite these severe consequences, the identified attacks exhibited substantially lower detectability across multiple estimator-level and navigation-state monitoring algorithms, with window-level detection AUROC values of approximately 0.50–0.56 and GRU-based episode-level alarm rates of 10–15\%. Furthermore, the analysis of signed temporal-difference errors revealed distinguishable distributional shifts at the episode level, suggesting that downstream value-function consistency may provide complementary evidence of adversarial effects that remain difficult to identify through conventional monitoring. Collectively, these findings demonstrate that statistical admissibility at the estimation layer does not necessarily guarantee the integrity of downstream autonomous decisions, highlighting the importance of incorporating policy sensitivity and mission-level consequences into adversarial edge-case assessment. Future work will investigate the generalizability of the proposed framework across different navigation policies and physical UAV platforms, alongside the development and experimental validation of cross-layer detection mechanisms that combine estimator-level consistency with downstream decision-making indicators. Further research will also be conducted to extend the adversarial edge-case-driven reliability assessment for multi-UAV systems in unmanned air traffic management (UTM) systems \cite{conrad2023intelligent, conrad2024simulating}. 
	
	\printbibliography

\end{document}